\documentclass[aps,prc,reprint,floatfix,superscriptaddress,nofootinbib]{revtex4-2}

\usepackage{amsmath}
\usepackage{amssymb}
\usepackage{graphicx}
\usepackage{bm}
\usepackage{color}
\usepackage{booktabs}
\usepackage{multirow}
\usepackage[section]{placeins}
\usepackage[colorlinks, linkcolor=red, anchorcolor=green,citecolor=blue ]{hyperref}
\begin{document}

\title{Radii and Scaling Relations of Dark-Matter-Admixed Neutron Stars}

\author{Zhi-yao Li}
\author{Wei-zhou Jiang}
\author{Dian Sun}
\affiliation{School of Physics, Southeast University, Nanjing 211189, China}

\begin{abstract}

The radius of a neutron star is closely related to the equation of state (EOS) of asymmetric nuclear matter, particularly to the nuclear symmetry energy. At fixed stellar mass, the neutron star (NS) radius and the nuclear-matter pressure near saturation density are known to satisfy an approximate pressure--radius ($R$-$P$) scaling relation, which reflects the relation between the geometry and matter within the strong interaction. In the presence of dark matter (DM) in NS, the modification to the geometry becomes a factor to change the $R$-$P$ scaling relation.  We investigate how GeV- and sub-GeV-scale DM modifies this relation in DM-admixed neutron stars (DANSs), within a  two-fluid framework bound by gravity only. We first extend the Buchdahl analytic solution to a two-fluid configuration to demonstrate the existence of such a scaling relation, and subsequently use a set of 21 nuclear EOSs to fit the  NM relation $R_{\rm NM}=\bar{C}P_{\rm NM}^{\eta}$ for  $1M_\odot$ and $1.4M_\odot$ DANSs. The exponent $\eta$ and coefficient $\bar{C}$ jointly quantify the DM-induced modification, depending on the DM candidate mass, the DM content of the star, and the self-interaction strength. The changes in $\eta$ and $\bar{C}$, relevant to the bosonic DM, occur mainly in the sub-GeV mass range, whereas the significant effects of fermionic DM persist up to the mass  approximately $1.5\,\mathrm{GeV}$. Increasing the DM content or strengthening a repulsive self-interaction generally increases $\eta$ and decreases $\bar{C}$. The modified $R$-$P$ relation therefore encodes information about the DM particle mass, its abundance in the star, and its self-interaction and affect the extraction of the nuclear EOSs through celestial observations. Interestingly, at DM candidate masses of several hundred MeV, both fermionic and bosonic DM can also produce DM-dominated low-mass ($1M_\odot$) DANSs with a tremendous  radial extension, for which the conventional scaling relation ceases to apply.
\end{abstract}


\maketitle

\section{Introduction}

With the advent of multimessenger astronomy, neutron stars (NS), as natural laboratories for testing the equation of state (EOS) of extremely dense nuclear matter, have attracted increasing attention in the astrophysics of compact objects and nuclear physics\cite{Lattimer2007, Abbott2018}. Determining the EOS of dense matter at extreme densities is  not only important for constraining the structure of the QCD phase diagram in the low-temperature, high-density regime, but is also essential for describing the structure, evolution, and mergers of neutron stars. The nuclear EOS encodes the pressure--energy density relation in NS interiors. Within the framework of general relativity, solving the Tolman--Oppenheimer--Volkoff (TOV) equations \cite{Oppenheimer1939} with this EOS as input yields the neutron star mass, radius, and internal matter distribution.
The microscopic description of NS matter nevertheless remains uncertain at two distinct levels. At the level of constituent degrees of freedom, dense matter may remain purely hadronic \cite{Akmal1998, Oertel2017}, undergo deconfinement to quark matter at hybrid or pure quark phases \cite{Glendenning1992q,Alford2008,Zhang2024}, or contain components beyond the standard model like dark matter (DM) components \cite{Sandin2009,Leung2011,Xiang2014}. Even for a specified composition, the in-medium many-body dynamics still relies on modelings, roughly categorized by microscopic and phenomenological approaches.  Microscopic approaches usually include Brueckner theory \cite{Day1967,Br1990,Song1998} and Green's-function methods \cite{Wiringa1988,Di2003}, often combined with nuclear forces from chiral effective field theory \cite{Fu1996,Fi2006,Hebeler2013}, while phenomenological descriptions include relativistic mean-field and energy-density-functional models \cite{Walecka1974, Dutra2014}. Consequently, EOSs derived from theory or constrained by experiments retain substantial uncertainties, especially at high density \cite{Lattimer2007,Li2008,Lattimer2012,Oertel2017}. NS matter is highly asymmetric matter in chemical equilibrium, and its EOS depends on both the isoscalar interaction and the density dependence of the symmetry energy. The stiffness of symmetric matter at high density is closely related to the maximum NS mass, whereas the symmetry energy near saturation density plays a key role in determining the stellar radius \cite{Lattimer2004,Lattimer2007}. EOS uncertainties also propagate to the composition, possible phase transitions \cite{Glendenning1992,Jiang2006}, and thermal evolution of neutron stars \cite{Yakovlev2004, Page2006}.

Connecting microscopic EOS properties with macroscopic NS observables allows for constraining the EOS of asymmetric nuclear  matter through celestial observations. In a static star, pressure generated by Fermi motion and nuclear interactions balances gravitational compression, as expressed by the TOV equations. Different EOSs therefore lead to different pressure profiles, metric gradients, and stellar properties.  Nevertheless, a number of approximately universal scaling relations have been identified either between observables and the EOS quantities or among different observables \cite{Andersson1998,Lattimer2001,Breu2016,Yagi2013_Science}. Results obtained with different EOSs approximately follow the same scaling curves, with only a weak dependence on the individual model in the sense of statistics.  Because NS observables encode EOS information, such relations can provide a practical route for constraining the nuclear EOS.

The existence of these scaling relations can be understood with analytic stellar models. By imposing specific matter profiles or metric functions, Tolman and Buchdahl obtained analytical solutions for compact-star interiors \cite{Buchdahl1959,Buchdahl1967,Tolman1939}. Their analytic metric, pressure, and energy-density profiles make the connection between the EOS and stellar observables explicit. Numerical TOV calculations then reconfirm these relations with realistic EOSs and determine their parameters more accurately. The existence of these scaling relations is generally because of the established balance between the geometric  gravity and the microscopic pressure of dense matter. The gravity responds mainly to the global mass distribution, while the mass--radius relation can be translated inversely into the nuclear EOS by virtue of the Lindblom method \cite{Lind1992,Chen2015}. In the past, many  scaling relations have been discovered, and examples include the relation between the gravitational-wave $f$-mode frequency and mean stellar density \cite{Andersson1998}, the nearly EOS-independent relation between the maximum masses of uniformly rotating and nonrotating neutron stars \cite{Breu2016}, the radius--pressure relation of Lattimer and Prakash \cite{Lattimer2001}, and the I-Love-Q relations among the moment of inertia, tidal deformability, and spin-induced quadrupole moment \cite{Yagi2013_Science}. These scaling relations establish the bridges universally between the characteristic quantities of the nuclear EOS and some NS  observables.

DM is the primary candidate for providing the additional gravitational attraction inferred from galactic rotation curves \cite{Mann2006} and accounts for approximately $26.8\%$ of the total cosmic energy density \cite{Planck2020}. Despite the gravitational evidence for DM, terrestrial experiments have yet to establish its particle nature or interactions. Over cosmic time, DM may accumulate in the strong gravitational field of a neutron star. Possible formation channels include gravitational capture of ambient DM \cite{Kouvaris2008,Brito2015}, mergers involving compact objects such as dark stars \cite{Diedrichs2023}, accretion of baryonic matter onto primordial dark objects \cite{Meliani2016}, and the conversion of Standard Model particles into dark-sector particles in dense stellar matter \cite{Baym2018, Berryman2022}. Meanwhile, multimessenger constraints on NS radii show that the tension in the results may exist in different observational channels. NICER measurements of PSR J0030+0451 \cite{Riley2019,Miller2019} and PSR J0740+6620 \cite{Riley2021,Miller2021} tend to favor relatively large radii, $R_{1.4}\gtrsim12.5$ km, and hence a larger symmetry-energy slope $L$ near saturation density. The tidal-deformability bound $\Lambda\leq590$ inferred from the binary-NS merger GW170817 instead favors more compact stars with smaller radii and a softer symmetry energy \cite{Abbott2018}. A DM component introduces additional energy density, pressure, and gravity while leaving the assumed nuclear EOS unchanged, thereby possibly reconciles the tension in the results extracted from different multimessenger observations  due to various sensitivities to new degrees of freedom in compact stars. Since the DM inclusion modifies the matter distribution, mass, and radius of NS,  significant DM signals may likely be revealed in such compact stars. These considerations motivate systematic studies of how different DM candidates and their EOSs affect compact-star structure \cite{Diedrichs2023, Goldman1989, Rutherford2023, Bertone2008, Ellis2018, Shirke2023_JCAP, Kark2022}.

Direct-detection experiments place stringent limits on the DM--nucleon scattering cross section, implying very weak momentum exchange between the two components on macroscopic scales \cite{PDG2024}. This validates an approximation in which normal matter (NM) and DM are coupled only through gravity \cite{Sandin2009, Tolos2015}. Under this scenario, the two components form distinct fluids following a two-fluid TOV formalism \cite{Cia2011,Xiang2014,jockel2024fermion}. A compact star, in which NM and DM coexist in the same spacetime, obey separate EOSs, and interact through their common gravitational field, is referred to here as a DM-admixed neutron star (DANS).

The existence of DM in compact stars can change  the relative spatial distributions of both fluids and therefore affect the hydrostatic structure at equilibrium  under the common gravitational field modified by the DM component \cite{Leung2011,Xiang2014, Tolos2015,Liebling2017}. The two-fluid calculations  using specific  DM models \cite{Xiang2014,Narain2006,Kark2022} further show that heavier DM candidates tend to be concentrated near the center, forming a compact dark core, whereas lighter DM candidates tend to spread over a larger region and can form a DM halo extending outside of the NM component \cite{Xiang2014,Ivan2020,Rout2025,Biesdorf2025}. These different distributions alter the mass and radius of the NM component and can consequently modify its scaling relations.

DANSs have attracted increasing attention over the past decade \cite{Xiang2014,Narain2006,Kark2022,Mukh2016,Moira2019,Das2020,Kain2021, Yang2021,Kark2022, Leung2022_Tidal,Collier2022,Komo2022,Miao2022,Deng2022, Shirke2023_JCAP, Shak2024,Sun2024, Hong2024, Liu2024, Yang2025,Zhou2026,Issi2026}. Given the influence of DM on the NM distribution, in this work we focus on the effects of fermionic and bosonic dark matter on the NM radius and the associated scaling relations, with particular emphasis on the differences in these effects. For normal neutron stars, Lattimer and Prakash found that the stellar radius scales approximately with the pressure of NS matter near saturation density as $R\propto P^{1/4}$ \cite{Lattimer2001}. The existence of this relation can be verified using the Buchdahl method \cite{Buchdahl1967}, which assumes a particular radial dependence of a metric function and yields  an analytic EOS of a special form. From this solution,  a power-law exponent is obtained close to $1/4$ for $1.4M_\odot$ NS \cite{Lattimer2001}. In reference to the accurately measured NS radius, the scaling relation can be well used to constrain the nuclear-matter pressure near saturation density. The DM component changes the spacetime metric and the gravitationally bound structure at equilibrium, and should therefore modify the corresponding pressure--radius relation. In this work, we study these modifications in DANSs for different DM scenarios. In the presence of dark matter, we first establish the existence of the radius–pressure scaling relation for a two-fluid configuration governed by the two-fluid TOV equations and then calculate its power-law exponent $\eta$ for different DM scenarios. We use $\eta$ and the corresponding scaling coefficient to describe how DM modifies the relation concerning nuclear matter stiffness and the NM radius of DANSs. The modified relation therefore links radius measurements to both the nuclear EOS and DM properties.

The remainder of this paper is organized as follows. Section II introduces the NM and DM EOSs, the two-fluid TOV equations, and the analytic derivation of the scaling relation. Section III presents numerical results for the DANS radii and the scaling relations under different DM scenarios, and discusses how these scenarios modify the scaling exponents and coefficients. Section IV summarizes our main conclusions.

\section{Theoretical Framework}
To investigate the matter distributions and macroscopic properties of DANSs and their dependence on nuclear models, we consider a broad set of nuclear EOSs and examine separately how fermionic and bosonic DM content in DANSs modify the $R-P$ scaling relation. For a combination of the specific nuclear and DM EOSs, the two-fluid TOV equations \cite{Xiang2014} are solved for static spherical DANS to obtain the radial distributions, masses, and radii of both fluids.
\subsection{Equations of State}

\subsubsection{Nuclear matter Equation of State}
To assess the model dependence from  various nuclear EOSs, we select 21 typical EOSs for cold, $\beta$-equilibrated NS matter. These EOSs belong to three categories according to their treatment of nuclear interactions. The first category consists of phenomenological relativistic density functionals: GM1 \cite{Glendenning1991}, PK1 \cite{Long2004}, TW99 \cite{Typel1999}, FSUGarnet \cite{ChenPiekarewicz2014,Chen2015}, SLC and SLCd \cite{Jiang2007_SLC,Jiang2012}, TM1w02 \cite{Ye2025}, DD2 \cite{Typel2010,Grill2014}, M1 and M3 \cite{Scurto2024}, MTVTC \cite{Maruyama2005}, and DDME-Y \cite{Taninah2020}. The second category comprises phenomenological nonrelativistic density functionals: BSK22 and BSK26 \cite{Goriely2013,Pearson2018}, and SLy5 \cite{Grams2022}. The third category includes microscopic models and models constrained by chiral symmetry: APR \cite{Akmal1998,DouchinHaensel2001}, WFF1 and WFF2 \cite{Wiringa1988}, BL \cite{Bombaci2018}, and H1 and H4 \cite{Grams2022}. Some of these EOSs are available directly from the CompOSE database \cite{Typel2015_CompOSE,Typel2022_CompOSE}. These typical  models cover a variety of in-medium nuclear interactions and yield a rather broad range of the NS mass--radius predictions. Note that all these models can produce the NS $2M_\odot$ constraint. The lepton sectors of all 21 $\beta$-equilibrated EOSs include both electrons and muons.

\begin{figure}[tbh]
    \centering
    \includegraphics[width=\columnwidth]{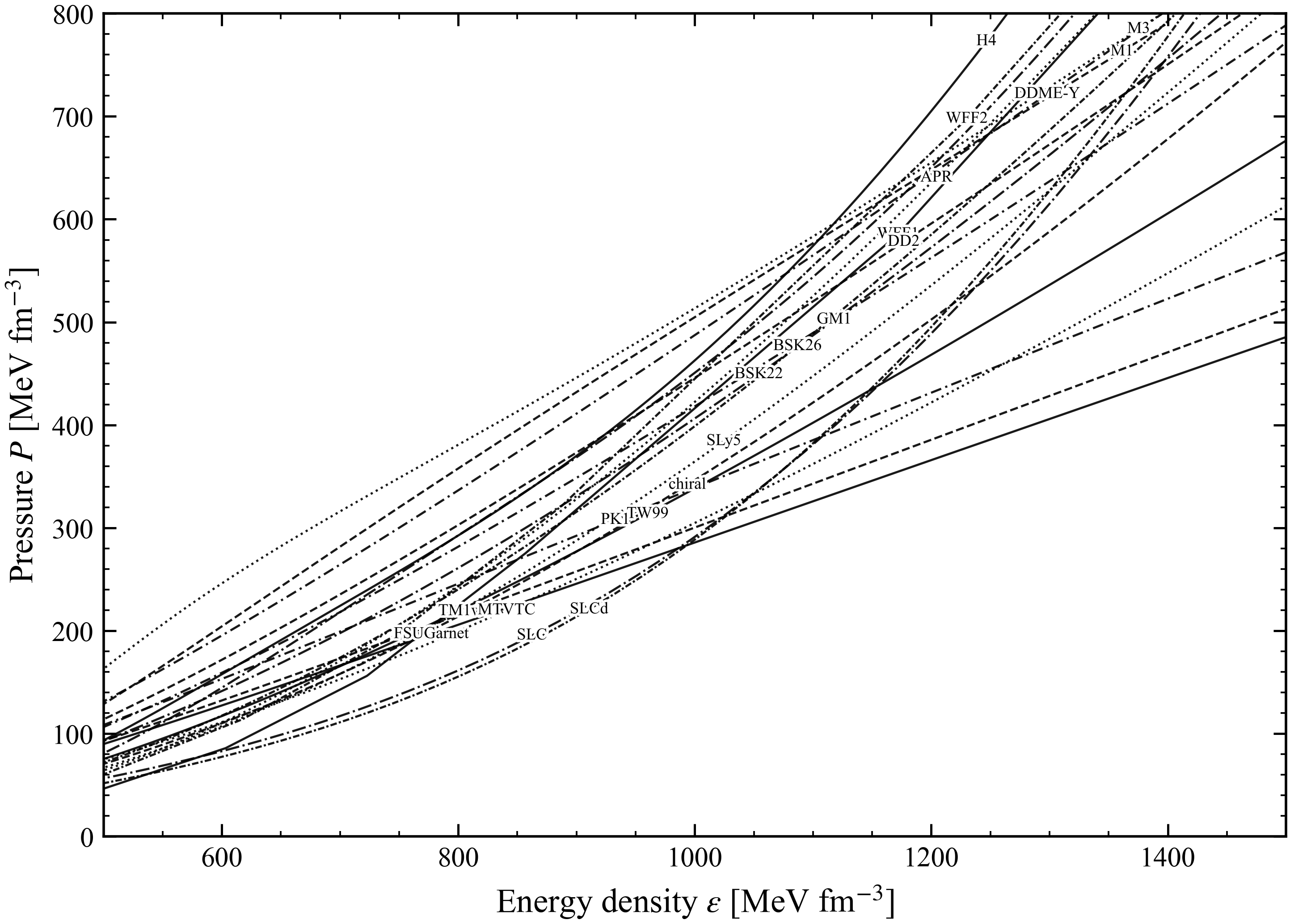}
    \caption{Equations of state for $\beta$-equilibrated NS matter with the selected nuclear models.  The curves show the pressure $P$ as a function of energy density $\varepsilon$ and illustrate the range of EOS stiffness covered by these models.}
    \label{fig:nm_eos}
\end{figure}

\begin{figure}[tbh]
    \centering
    \includegraphics[width=\columnwidth]{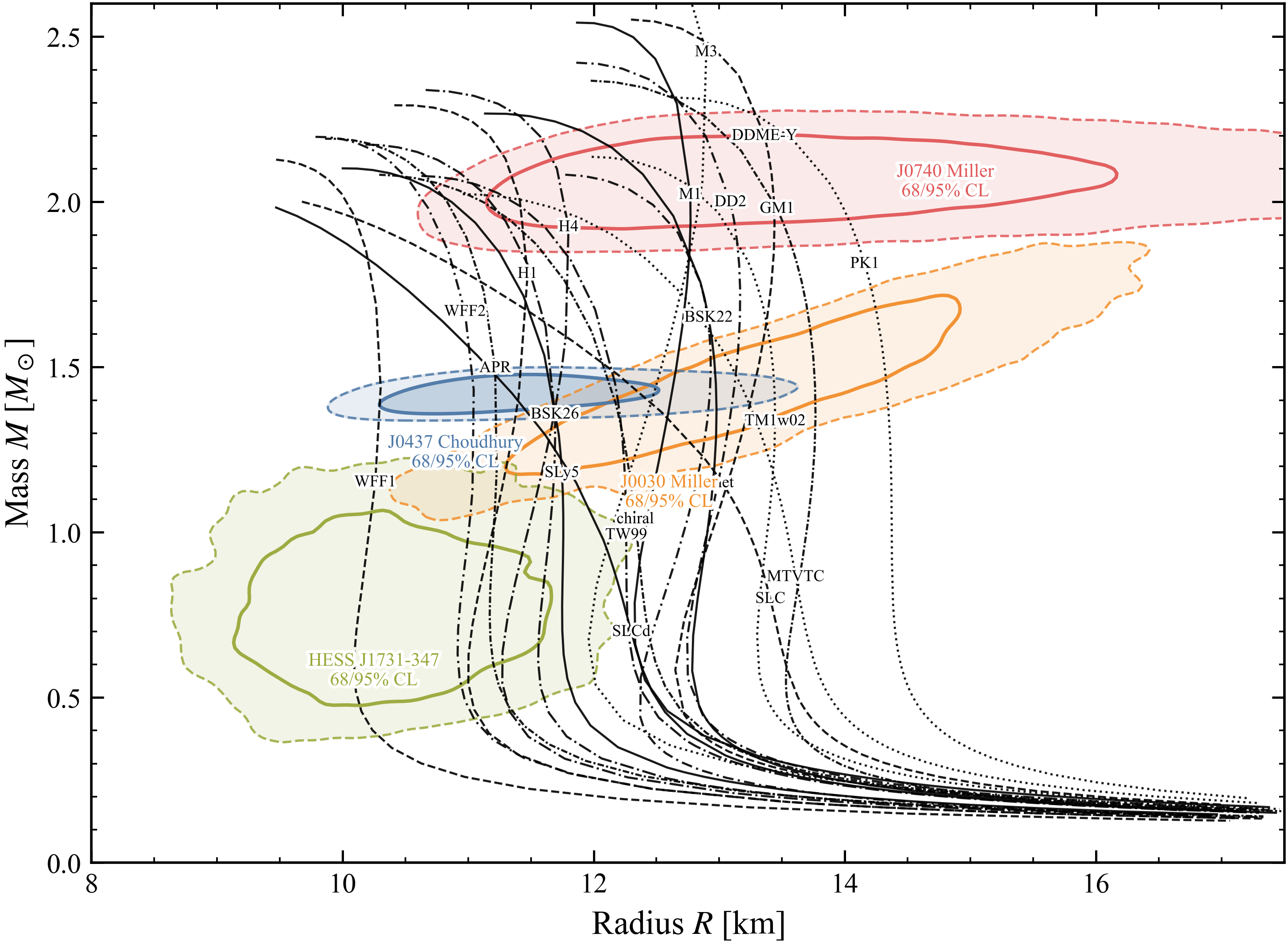}
    \caption{Mass--radius trajectories of normal NSs with various nuclear models. The solid and dashed contours with shaded areas represent the $68\%$ and $95\%$ confidence regions inferred from the NS observations, respectively. The black curves show the mass--radius relations obtained with the nuclear-matter EOSs used in this work.}
    \label{fig:MR_Summary_Single}
\end{figure}

Figures~\ref{fig:nm_eos} and \ref{fig:MR_Summary_Single} show, respectively, the EOSs of $\beta$-equilibrated NS matter and corresponding NS mass--radius trajectories with these models. Figure~\ref{fig:MR_Summary_Single} also displays some observational constraints in the mass--radius plane, including the NICER posteriors for PSR J0030+0451 \cite{Riley2019,Miller2019}, PSR J0740+6620 \cite{Riley2021,Miller2021}, and PSR J0437-4715 \cite{Choudhury2024_J0437}, together with the constraints associated with HESS J1731-347 \cite{Doroshenko2022_HESS,Doroshenko2022_HESS_Zenodo}. The shaded areas with solid and dashed contour lines represent the   $68\%$ and $95\%$ confidence regions from the NS observations, respectively.

From the surface inward, a neutron star is commonly divided into an outer crust, an inner crust, and a uniform liquid core. The outer crust consists of a lattice of nuclei immersed in an electron gas. Above the neutron-drip density, free neutrons appear and form the inner crust; at still higher densities, the matter becomes an approximately uniform, strongly interacting nuclear fluid. For each core EOS, we determine the crust--core boundary either using the transition densities from the database \cite{Typel2015_CompOSE,Typel2022_CompOSE} or with the thermodynamic stability condition \cite{Lattimer2007}. Below the crust-core transition density, piecewise EOSs are considered in various density regions \cite{BPS1971,Iida1997}. We use the FMT model \cite{BPS1971} in the dilute outermost region, $4.73\times10^{-15}\,\mathrm{fm}^{-3}\leq n_b\leq6.93\times10^{-13}\,\mathrm{fm}^{-3}$; the BPS model \cite{BPS1971} for the main part of the outer crust, $6.93\times10^{-13}\,\mathrm{fm}^{-3}<n_b\leq2.46\times10^{-4}\,\mathrm{fm}^{-3}$; and the multiphase-equilibrium EOS of Iida and Sato, based on a liquid-drop model \cite{Iida1997}, for the inner crust, $2.46\times10^{-4}\,\mathrm{fm}^{-3}\leq n_b\leq n_t$. This common layered construction allows for the  TOV integration to proceed  from the central core to the NM surface while maintaining a consistent low-density treatment for all core EOSs.
\subsubsection{Fermionic DM Equation of State}
Little information about the DM property including whether the DM has self-interactions is confirmed by terrestrial experiments.
For a possible fermionic DM component in NSs, its EOS relies on the DM scenario one takes. Here,  we adopt a self-interacting fermionic model analogous to relativistic mean-field theory \cite{Narain2006,Xiang2014}. The DM particles interact through the exchange of the attractive scalar ($\phi$) and repulsive vector $V_\mu$ mediator fields. The corresponding Lagrangian density is

\begin{equation}
\begin{split}
\mathcal{L}_D &= \bar{\psi}_D \left[ \gamma_\mu(i\partial^\mu - g_v V^\mu) - (M_D - g_s \phi) \right]\psi_D \\
&\quad + \frac{1}{2}(\partial_\mu\phi\partial^\mu\phi - m_s^2\phi^2) \\
&\quad - \frac{1}{4}D_{\mu\nu}D^{\mu\nu} + \frac{1}{2}m_v^2 V_\mu V^\mu
\end{split}
\label{eq:FermionL}
\end{equation}
Here, $D_{\mu\nu}=\partial_\mu V_\nu-\partial_\nu V_\mu$ is the field-strength tensor of the vector mediator; $g_s$ and $g_v$ are in turn the scalar and vector coupling constants, $M_D$ is the bare DM particle mass, and $m_s$ and $m_v$ are the corresponding mediator masses.
Within the mean-field approximation, it is useful to define the parameters $C_s$ and $C_v$:
\begin{equation}
C_s=\frac{g_s}{m_s},\qquad C_v=\frac{g_v}{m_v}.
\label{eq:CsCvDefinition}
\end{equation}
The DM energy density $\varepsilon_D$ and pressure $p_D$ are obtained from the zero-temperature Fermi-sea integrals and the mean-field contributions of the mediators:
\begin{equation}
\begin{split}
\varepsilon_D &= \frac{2}{(2\pi)^3}\int^{k_{F_D}} d^3k \sqrt{k^2 + (M_D^*)^2} \\
&\quad + \frac{C_v^2}{2}\rho_D^2 + \frac{1}{2C_s^2}(M_D - M_D^*)^2,
\end{split}
\end{equation}

\begin{equation}
\begin{split}
p_D &= \frac{1}{3}\frac{2}{(2\pi)^3}\int^{k_{F_D}} d^3k \frac{k^2}{\sqrt{k^2 + (M_D^*)^2}} \\
&\quad + \frac{C_v^2}{2}\rho_D^2 - \frac{1}{2C_s^2}(M_D - M_D^*)^2.
\end{split}
\end{equation}
The effective DM mass is defined as $M_D^*=M_D-g_s\phi_0$, while $\rho_D$ and $k_{F_D}$ denote the DM number density and Fermi momentum, respectively. In the numerical calculation, $M_D^*$ is solved self-consistently from the mean-field equation at each $\rho_D$ or $k_{F_D}$, and the DM EOS is then obtained by calculating $\varepsilon_D$ and $p_D$.

The DM component is included in the DANS  within the framework of two-fluid TOV equations \cite{Sandin2009,Cia2011,Xiang2014}, in which only pure gravitational interaction exists between the NM and DM. Recent direct-detection experiments such as XENONnT and LUX-ZEPLIN continue to tighten the limits on the DM--nucleon scattering cross section \cite{Aprile2023_XENONnT,Aalbers2022_LZ}. The two-fluid framework can even be applicable for  describing the limiting case in which nongravitational interactions are very weak. In reference to  previous studies of DANSs with GeV- and sub-GeV-scale fermionic DM \cite{Narain2006,Cia2011,Xiang2014,jockel2024fermion,Kark2022}, the mass range of DM candidates we choose  is  as the following
\[
500\,\mathrm{MeV}\leq m_\chi\leq2000\,\mathrm{MeV}.
\]
We will examine how the mass of DM candidate affects the DM distribution inside the star and the resulting DANS structure.

Once the DM particle mass is specified, the mean-field contribution to the DM EOS is solely expressed by the parameters $C_s$ and $C_v$.  For simplicity, the masses of scalar and vector mediators are set to be $m_s=m_v=1\,\mathrm{GeV}$. We consider three cases for the DM self-interactions: scalar and vector interactions,  separately, and both of them simultaneously. The coupling ratios are treated as free parameters within the ranges
\begin{equation}
0\leq C_s\leq5\,\mathrm{GeV}^{-1},\qquad
0\leq C_v\leq8\,\mathrm{GeV}^{-1}.
\end{equation}

\subsubsection{Bosonic DM Equation of State}

For bosonic DM, we adopt the repulsively self-interacting model introduced by Karkevandi et al. \cite{Kark2022}. The DM candidate is described by a complex scalar field $\phi$ of mass $m_\chi$, with a quartic self-interaction potential $V(\phi)=\lambda|\phi|^4/4$, where $\lambda$ is a dimensionless self-coupling constant. We consider a zero-temperature Bose--Einstein condensate for bosonic DM and use $\lambda>0$ as a free parameter adjusting the strength of the repulsive self-interaction.

The relation between the pressure $P$ and energy density $\varepsilon$ of this bosonic DM system is analytically expressed as
\begin{equation}
P = \frac{m_{\chi}^4}{9\lambda} \left( \sqrt{1 + \frac{3\lambda}{m_{\chi}^4} \varepsilon} - 1 \right)^2.
\end{equation}
Here, the DM EOS is jointly determined by the particle mass $m_\chi$ and the dimensionless coupling $\lambda$. The mass of bosonic DM candidate is taken to  cover the sub-GeV to GeV regime \cite{Kark2022,Bringmann2019_CRDM}: $300\,\mathrm{MeV}\leq m_\chi\leq2000\,\mathrm{MeV}$ in studying the dependence on the bosonic particle mass. At fixed $m_\chi$, increasing $\lambda$ raises the pressure at a given energy density. We use $\lambda=\pi$ for the mass scan and later examine how stronger repulsion affects the scaling relation by comparing the results with $\lambda=\pi$ and $4\pi$.

\begin{figure}[tbh]
    \centering
    \includegraphics[width=\columnwidth]{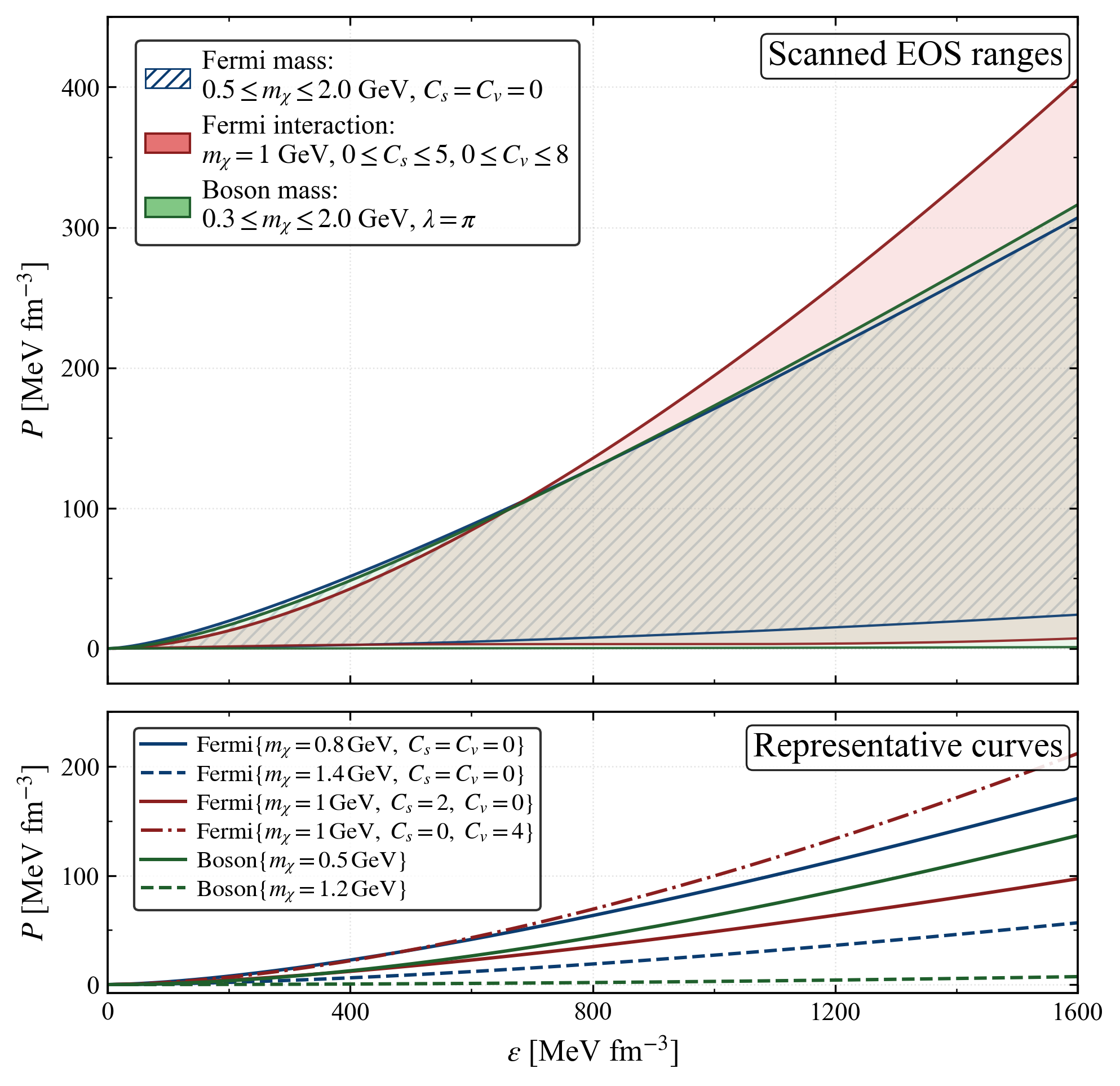}
    \caption{ DM EOSs with parameter scans and representative curves. The shaded regions in the upper panel show the pressure ranges covered by the fermionic mass scan ($0.5\leq m_\chi\leq2.0\,\mathrm{GeV}$, $C_s=C_v=0$), the fermionic self-interaction scan ($m_\chi=m_s=m_v=1\,\mathrm{GeV}$, $0\leq C_s\leq5\,\mathrm{GeV}^{-1}$, and $0\leq C_v\leq8\,\mathrm{GeV}^{-1}$), and the bosonic mass scan ($0.3\leq m_\chi\leq2.0\,\mathrm{GeV}$, $\lambda=\pi$). The lower panel presents representative EOSs from each scan.}
    \label{fig:dm_eos}
\end{figure}

Figure~\ref{fig:dm_eos} summarizes the DM EOSs over the parameter ranges considered above. The results are grouped into the mass scan and  self-interaction scan for the fermionic DM candidate, and the mass scan for bosonic DM candidate. The upper panel shows the overall pressure ranges spanned by the three EOS families, while the lower panel displays representative curves illustrating the changes associated with varying the particle mass or interaction strength. Increasing the fermionic DM mass or introducing scalar attraction softens the EOS, whereas decreasing the mass or strengthening vector repulsion raises the pressure. The bosonic EOSs   with $\lambda=\pi$ are also shown for comparison with the fermionic models.

\subsection{Two-Fluid TOV Equations}
To quantify the effects of DM on the macroscopic properties of neutron stars, we describe the internal structure of a DANS using a two-fluid model. DM and NM are assumed to interact only through gravity \cite{Sandin2009,Xiang2014}. The system can therefore be treated as two independently conserved fluids at hydrostatic equilibrium within a common gravitational field. In units of $G=c=1$, the event interval in  the static, spherically symmetric Schwarzschild metric is given as
\begin{equation}
\begin{aligned}
ds^2={}&-e^{2\nu(r)}dt^2
+\left[1-\frac{2M(r)}{r}\right]^{-1}dr^2\\
&+r^2\left(d\theta^2+\sin^2\theta\,d\phi^2\right).
\end{aligned}
\label{eq:schwarzschild_metric}
\end{equation}
Within this Schwarzschild metric, the two-fluid TOV equations take the form \cite{Xiang2014}:
\begin{equation}
\begin{aligned}
\frac{d P_{i}(r)}{d r} =-\frac{\left[\varepsilon_{i}(r)+P_{i}(r)\right]\left[4 \pi r^{3} P(r)+M(r)\right]}{r[r-2 M(r)]},
\end{aligned}
\label{eq:dual_tov}
\end{equation}

\begin{equation}
\frac{d M_{i}(r)}{d r} =4 \pi r^{2} \varepsilon_{i}(r).
\end{equation}
Here, the index $i$ represents the NM or DM component, and $M_i(r)$ is the corresponding  mass enclosed within radius $r$.

The total mass $M(r)$ and total pressure $P(r)$ are the sums of the two contributions:
\begin{equation}
M(r) = M_{\text{NM}}(r) + M_{\text{DM}}(r), \quad P(r) = P_{\text{NM}}(r) + P_{\text{DM}}(r).
\end{equation}
Although the two components do not interact within the standard model, the gravity from DM changes the metric-potential gradient $d\nu/dr$ and consequently modifies the NM distribution.
\subsection{Analytic Scaling Relation}

In this subsection, we  demonstrate how a pressure--radius scaling relation arises in a two-fluid system, with particular emphasis on the Buchdahl solution. Numerical solution to the TOV equations determines the stellar structure for a specific EOS, whereas analytic expressions help identify how DM modifies the relation through the total pressure, matter distribution, and compactness. Buchdahl assumed a particular radial dependence of the angular metric component and obtained the following analytic EOS according to the matter boundary conditions \cite{Buchdahl1967}:
 \begin{equation}
    \varepsilon    = 12\sqrt{p_*P}-5P.
    \label{eq:eos}
\end{equation}
Here, $p_*$ is a parameter related to the central pressure and energy density. Using the Buchdahl EOS,  the following relation at fixed stellar mass and a specified reference density can be derived \cite{Lattimer2001}:
\begin{equation}
   \eta= \left. \frac{d\ln R}{d\ln P} \right|_{\varepsilon,M}
    =
    \frac{1}{2}
    \left( 1 - \frac{5}{6}\sqrt{\frac{P}{p_{*}}} \right)
    \frac{(1-\beta)(1-2\beta)}{1-3\beta+3\beta^2},
    \label{eq:Lattimer_scaling1}
\end{equation}
where $\beta=GM/Rc^2$ is the stellar compactness. Substituting NS representative values, as discussed below, gives approximately the scaling exponent $\eta=0.31$, demonstrating the existence of a scaling relation. Numerical extraction from a collection of selected EOSs further yields  a somehow refined form $R\propto P^{1/4}$. This  power-law scaling relation directly connects the feature of nuclear EOS with the macroscopic NS radius.

Hereafter, the power-law relation between the NM radius $R_{\rm NM}$ at fixed stellar mass and the NM pressure $P_{\rm NM}$ evaluated at the representative baryon densities $\rho_0$, $1.5\rho_0$, and $2\rho_0$ is referred to as the pressure--radius ( $R$-$P$), scaling relation. In the numerical calculations, we consider DANSs with total masses of $1.0M_\odot$ and $1.4M_\odot$ and determine the exponent $\eta$ in the scaling relation $R=\bar{C} P^\eta$.

To test the robustness of this $R$-$P$ relation against the choice of analytic metric ansatz, we also derive analogous relations from the Tolman VII and Finch--Skea solutions \cite{Tolman1939,FinchSkea1989}. For the representative values used in the analysis by Lattimer and Prakash \cite{Lattimer2001}, $M=1.4M_\odot$, $R=15\,\mathrm{km}$, and $\varepsilon\simeq m_b\rho_0\simeq2.02\times10^{-4}\,\mathrm{km}^{-2}$, the resulting exponents are $0.3124$ for the Buchdahl solution, $0.4040$ for the Tolman VII solution, and $0.2212$ for the Finch--Skea solution. Current constraints from the tidal deformability of GW170817 \cite{Abbott2018} and NICER measurements of pulsars \cite{Riley2019, Miller2021} favor a representative radius near $12\,\mathrm{km}$ for a $1.4M_\odot$ neutron star \cite{ji2025equation}. For the updated choice $M=1.4M_\odot$ and $R=12\,\mathrm{km}$, the corresponding exponents become $0.3825$, $0.4601$, and $-0.0348$ for the Buchdahl, Tolman VII, and Finch--Skea solutions, respectively.

As the compactness increases, the exponent obtained from the Finch--Skea solution departs substantially from the range expected from current observations. The Buchdahl and Tolman VII results remain more stable and physically plausible. In the following analytic treatment of a DM fluid, we therefore adopt the Buchdahl metric, whose result is also closer to that of the numerical calculation. Appendix~\ref{app:buchdahl_scaling} shows how the DM-induced change in the NM distribution modifies the analytic energy-density--pressure relation. The resulting EOS for the total fluid in a DANS is similarly given as
\begin{equation}
    \varepsilon_t = 12\sqrt{p_{*,t}(P_{\rm NM}+P_{\rm DM})} - 5(P_{\rm NM}+P_{\rm DM})
    \label{eq:total_eos}
\end{equation}
The scaling exponent is modified accordingly. We define the following dimensionless  $\eta$ for visible matter:
\begin{equation}
\begin{aligned}
\eta
&\equiv
\left.
\frac{d\ln R_{\rm NM}}{d\ln P_{\rm NM}}
\right|_{\varepsilon_t,M}  \\
&=
\frac{1}{2}
\frac{P_{\rm NM}}{P_{\rm NM}+P_{\rm DM}}
\left( 1 - \frac{5}{6}\sqrt{\frac{P_{\rm NM}+P_{\rm DM}}{p_{*,t}}} \right) \\
&\quad\times
\frac{(1-\gamma\beta)(1-2\gamma\beta)}
{1-3\gamma\beta+3\gamma^2\beta^2}.
\end{aligned}
\label{eq:our_scaling}
\end{equation}
In Eq.~\eqref{eq:our_scaling}, $p_{*,t}$ is a constant that characterizes the pressure scale of the total fluid; further details are given in Appendix~\ref{app:buchdahl_scaling}. To isolate the explicit dependence on the NM pressure in this analytic estimate, we evaluate the partial derivative at fixed $P_{\rm DM}$, such that $dP_t=dP_{\rm NM}$. The corresponding logarithmic derivative yields the factor $P_{\rm NM}/P_t$. The observable radius of a DANS is the NM radius $R_{\rm NM}$, and the reference pressure is likewise the NM pressure $P_{\rm NM}$ at the selected density. The parameter $\gamma=R_{\rm NM}/R_{\rm total}$ describes the radial expansion of the NM component relative to the outer boundary of the total fluid. Here, $R_{\rm total}$ is the outermost boundary of the two-fluid system and can exceed $R_{\rm NM}$ when DM forms an extended halo. Thus, $0<\gamma\leq1$, with $\gamma=1$ when DM lies entirely within the NM radius. The compactness defined using the NM radius is $\beta=GM/R_{\rm NM}c^2$, while $\gamma\beta$ in the analytic expression is the effective compactness associated with the total-fluid boundary. In Eq.(\ref{eq:our_scaling}), $\varepsilon_t$ and $P_t$ denote the total energy density and pressure of the combined NM and DM fluids at the reference density. When the DM component vanishes, $\varepsilon_t$ and $P_t$ reduce to the nuclear-matter quantities $\varepsilon$ and $P$, and $\gamma=1$; Eq.~\eqref{eq:our_scaling} then reduces to the single-fluid form.

Comparison with Eq.~\eqref{eq:Lattimer_scaling1} shows that the scaling exponent of a DANS is no longer determined solely by the compactness $\beta$. The static structure is established upon the balance between gravitational attraction and pressure resistance, while the gravity provided by the DM alters the distribution and  boundary of the NM component. We characterize  this change through $\gamma=R_{\rm NM}/R_{\rm total}$ and the effective compactness $\gamma\beta$. At fixed DANS mass, different DM contents produce different values of $\gamma$ and $\gamma\beta$. The modification of the $R$-$P$ relation therefore depends mainly on the spatial distribution and total amount of DM, both of which, as shown below, depend strongly on the mass of the  DM candidate.

From the perspective of laboratory and observational constraints, the $R$-$P$ scaling relation provides a useful connection between nuclear matter properties and macroscopic NS observables \cite{Lattimer2001,Lattimer2007}. For NSs without DM, NICER radius measurements and tidal-deformability constraints from binary NS mergers restrict the radius at a given mass \cite{Riley2019,Miller2019,Riley2021,Miller2021,Abbott2018}. Heavy-ion collisions, electron scattering, and nuclear structure experiments, together with nuclear many-body calculations, well constrain the nuclear EOS between approximately $\rho_0$ and $2\rho_0$, although the density dependence of the symmetry energy remains uncertain \cite{Lattimer2007,Li2008,Lattimer2012, Hebeler2013, Oertel2017,Baldo2016,Li2018}. This uncertainty directly propagates to the pressure of NS matter. Through the $R$-$P$ scaling relation of the pure NS, an accurately measured NS radius can constrain the pressure and symmetry energy of asymmetric nuclear matter \cite{Lattimer2001}. Conversely, sufficiently precise laboratory constraints on the EOS can be translated into a prediction for the NS radius. In other words, the relation  provides a test of whether a nuclear EOS is compatible with an observed radius. In the presence of DM, this test becomes particularly significant, since the DM can modify the scaling relation, see Eq.~\eqref{eq:our_scaling}. If both the nuclear-matter pressure and NS radius become tightly constrained, deviations from the scaling relation based on  purely nuclear-matter EOSs can be used to extract the information of the DM particle mass, content, and self-interaction. Since Eq.~\eqref{eq:our_scaling} is built on the quite special metric and EOS, more reliable and  accurate  scaling relations should be extracted numerically from a collection of realistic nuclear models to assess the effects of various  DM scenarios, as done below.
\section{Results and Discussion}

The analytic results above show that the $R$-$P$ scaling relation for a DANS is not only determined by the nuclear EOS but also affected by the gravitational field from the DM. A similar physical picture also appears in  previous studies: a DM component can systematically modify the mass-radius relation, tidal deformability, and related universal relations of a DANS \cite{Grippa2025}. For example, Leung et al. studied the tidal Love numbers and tidal deformabilities of DANSs in a two-fluid framework and found departures from the corresponding relations for normal neutron stars \cite{Leung2022_Tidal}. Kumar and Sotani further combined mass-radius and tidal-deformability calculations with NICER and GW170817 constraints to restrict the DANS parameter space, showing that DM can modify the compactness--tidal-deformability universal relation \cite{Kumar2024}. These results indicate that DM can affect not only individual macroscopic observables but also correlations among them. Here we examine how different DM scenarios modify the $R$-$P$ relation between the nuclear-matter pressure and the NM radius. Such modifications directly affect astrophysical inferences of the nuclear EOS and provide a reference for searches for DM signatures in compact stars.

In numerical calculations, we select 21 representative nuclear EOSs from different treatments of many-body interactions and introduce either fermionic DM or self-repulsive bosonic DM. The configurations of the two fluids in their common gravitational field are obtained by solving the two-fluid TOV equations. Since the two components are coupled only through gravity, the surface of each component is defined by the radius at which its pressure vanishes. The NM radius $R_{\rm NM}$ and DM radius $R_{\rm DM}$ therefore generally differ, and $R_{\rm NM}$ defines the electromagnetically observable surface. Below, we vary the mass of the DM candidate $m_\chi$, the central DM density ratio over NM $\zeta_c(=\varepsilon_{DM,c}/\varepsilon_{NM,c})$, and the self-interaction strength, and compare the resulting changes in the scaling exponent $\eta$ and coefficient $\bar{C}(\rho_k)$ in the scaling relation.

\subsection{Effects of $m_\chi$ on the Matter Distribution and Scaling Relation}

\begin{figure}[tbh]    \centering
    \includegraphics[width=\columnwidth]{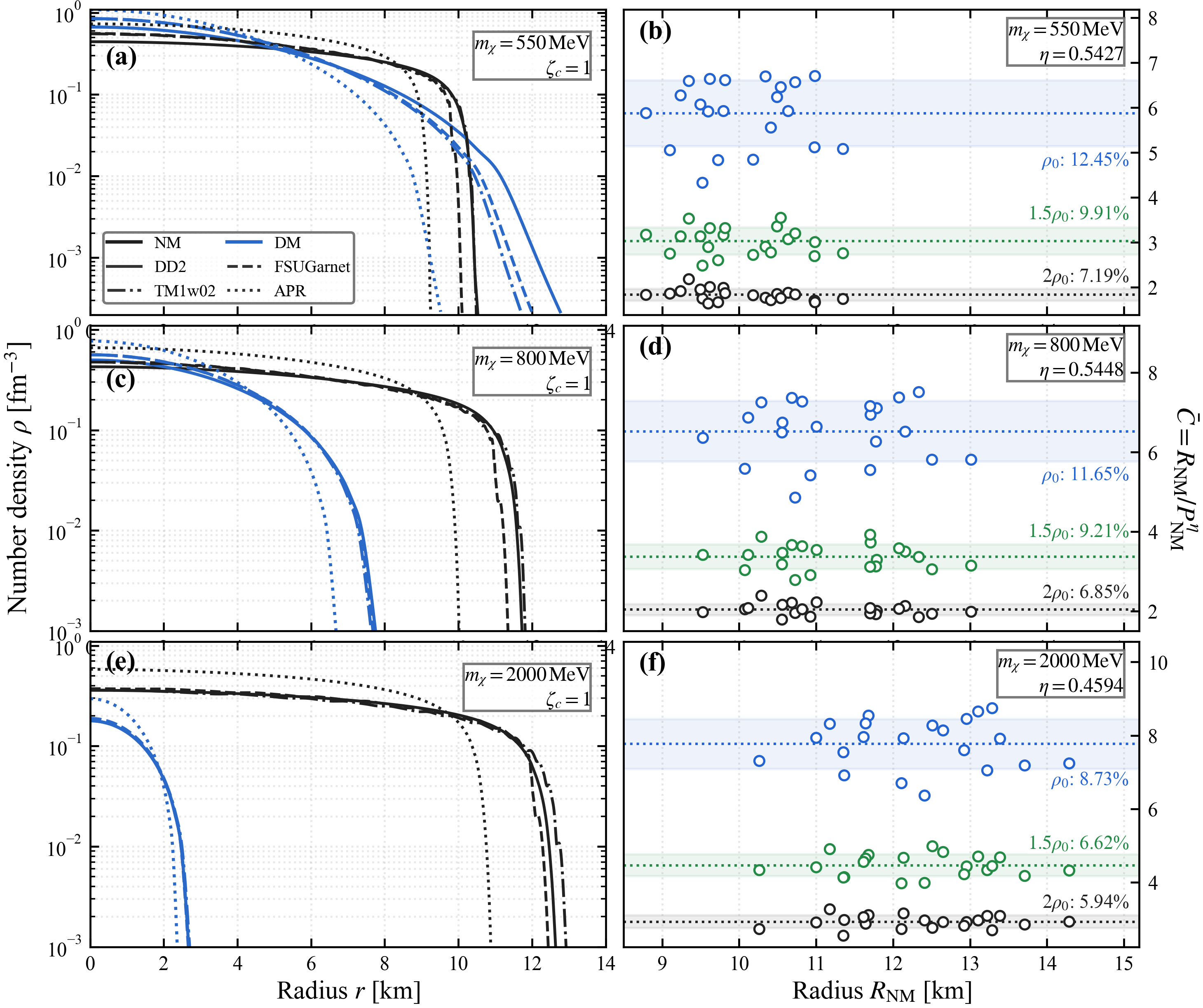}
    \caption{DM distributions and $R$-$P$ scaling relations for $1.4M_\odot$ DANSs. From top to bottom, the panels correspond to the cases of $m_\chi=550$, $800$, and $2000\,\mathrm{MeV}$, respectively. The left column shows the number-density profiles for four representative nuclear EOSs. The black and blue curves denote NM and DM, respectively, while the solid, dashed, dash-dotted, and dotted curves stand for the results with DD2, FSUGarnet, TM1w02, and APR, respectively. The right column shows the scaling fits obtained from the 21 nuclear EOSs for $m_\chi=550$, $800$, and $2000\,\mathrm{MeV}$, with fitted exponents $\eta=0.5427$, $0.5448$, and $0.4594$, respectively. The dashed lines denote the sample expectation value $\bar{C}$ of $C_i=R_{\rm NM}/P_{\rm NM}^{\eta}$, the shaded regions indicate $\bar{C}\pm\sigma$, and the relative scatter $\sigma/\bar{C}$ is given in the lower right of each panel.}
    \label{fig:ds_1p4}
\end{figure}

To illustrate how the mass of the DM candidate $m_\chi$ affects the radial distribution of DM and, in turn, the macroscopic scaling relation, we select four representative nuclear EOSs: DD2, FSUGarnet, TM1w02, and APR. We calculate various  $1.4M_\odot$ DANS configurations with $m_\chi=550$, $800$, and $2000\,\mathrm{MeV}$ at a fixed central energy-density ratio $\zeta_c\equiv\varepsilon_{\mathrm{DM,c}}/\varepsilon_{\mathrm{NM,c}}=1$. As shown in the left column of Fig.~\ref{fig:ds_1p4}, the $m_\chi$ value may actually determine  the radial extent of the dark component. A larger degeneracy pressure of the lightest DM candidate ($550\,\mathrm{MeV}$) produces an extended dark halo surrounding the NM component (Dark Halo, $\gamma<1$). The $800\,\mathrm{MeV}$ case lies in the transition region between an extended halo and a compact core. For the heaviest particles ($2000\,\mathrm{MeV}$), the lower degeneracy pressure allows for gravitational binding to dominate, producing a compact dark core embedded within the NM component (Dark Core, $\gamma=1$).

To extract the scaling parameters quantitatively, we use linear least-squares regression. At a given fiducial density $\rho_k$, we define
\begin{equation}
C_i(\rho_k,\eta)=\frac{R_{{\rm NM},i}}{\left. P_{{\rm NM},i}^{\eta} \right |_{\rho_k}},
\end{equation}
where the index $i$ labels the nuclear EOS. Lattimer and Prakash identified   $R$-$P$ scaling relations for neutron stars without DM over the density range $\rho_0$--$2\rho_0$ \cite{Lattimer2001,Lattimer2007}. We similarly consider three fiducial densities near saturation: $\rho_k=\rho_0$, $1.5\rho_0$, and $2\rho_0$. In Refs.~\cite{Lattimer2001,Lattimer2007}, the scaling exponent was estimated rather than optimized by fitting. Here, at each fiducial density, we fit the scaling parameters  from the 21 EOSs using ordinary least squares:
\begin{equation}
C_i=a_k R_{{\rm NM},i}+b_k .
\end{equation}
The least-squares procedure determines the best-fit line at each density. If the EOS sample obeys $R_{\rm NM}\propto P_{\rm NM}^{\eta}$, then $C_i$ should be approximately constant and the corresponding slope $a_k$ should approach zero. For each trial value of $\eta$, we obtain $a_k(\eta)$ at $\rho_0$, $1.5\rho_0$, and $2\rho_0$, and select the value of $\eta$ that minimizes $\sum_k a_k^2(\eta)$. The fitted slopes are small. For example, for the $1.4M_\odot$ DANSs shown in the right column of Fig.~\ref{fig:ds_1p4}, the absolute slopes for $m_\chi=550$, $800$, and $2000\,\mathrm{MeV}$ are all of order $10^{-2}$, with a maximum of approximately $8\times10^{-2}$.

Once the common exponent $\eta$ has been determined, the scaling coefficient at each fiducial density is defined through the scaling relation:
\begin{equation}
R_{\rm NM}=\bar{C}(\rho_k)P_{\rm NM}(\rho_k)^{\eta},
\qquad
\rho_k=\rho_0,\ 1.5\rho_0,\ 2\rho_0,
\label{eq:three_C_scaling}
\end{equation}
where $\bar{C}(\rho_k)$ is the sample expectation value of $C_i$ over the 21 EOSs at density $\rho_k$.

The right column of Fig.~\ref{fig:ds_1p4} shows the results of the scaling relation for $m_\chi=550$, $800$, and $2000\,\mathrm{MeV}$. The shaded intervals $\bar{C}\pm\sigma$ and the relative scatter $\sigma/\bar{C}$ show that the NM radius remains well correlated with the nuclear-matter pressure at the fiducial density. As $m_\chi$ decreases, the points in the right column shift toward smaller $R_{\rm NM}$, which is in nice agreement with the shift in the NM distribution, as shown on the left column of  Fig.~\ref{fig:ds_1p4}. These  changes in  $R_{\rm NM}$ and scaling parameters versus $m_\chi$ are consistent with those of the structural quantities listed in Table~\ref{tab:mass_scaling_full}, in which the properties of a single DANS are carried out with  the RMF model DD2. The decrease of  $R_{\rm NM}$ is mainly ascribed to the rise of the DM fraction at the smaller $m_\chi$ and  the corresponding reduction of the NM fraction in a DANS with a fixed total mass. The DM with the lighter candidate mass has a more extended distribution and consequently  contributes a larger mass fraction by integration over a larger spherical radius. In this regime, $R_{\rm NM}$ is governed not only by the nuclear EOS but also by the DM mass distribution and the partition ratio between the two fluids. Differences among the EOS pressures at the fiducial density persist, but the response of $R_{\rm NM}$ to those differences is modified by the DM content. Consequently, the relative scatter of $C_i=R_{\rm NM}/P_{\rm NM}^{\eta}$ increases as $m_\chi$ decreases. The scatter at $2\rho_0$ is notably smaller than that at $\rho_0$. Near saturation density, the pressure of NS matter is controlled primarily by the slope of the symmetry energy, which varies appreciably among the models and therefore produces a larger scatter at $\rho_0$. At higher densities, the pressure also depends on bulk nuclear-matter properties such as the incompressibility and skewness parameters:
\begin{equation}
\begin{aligned}
P(\rho)={}&\left(\frac{\rho}{\rho_0}\right)^2
\left[\frac{K_0}{9}(\rho-\rho_0)
-\frac{J_0}{54\rho_0}(\rho-\rho_0)^2\right]\\
&+\frac{\rho}{3}L(\rho)\alpha^2
+\rho E_{\rm sym}(\rho)x_p\alpha .
\end{aligned}
\end{equation}
Here, $K_0$, $J_0$, and $L(\rho)$ are in turn the incompressibility, skewness, and the slope of the symmetry energy, $\alpha\equiv(\rho_n-\rho_p)/\rho$ is the isospin asymmetry, and the last term is the lepton contribution with $x_p\equiv\rho_p/\rho$ being the proton fraction. At $2\rho_0$, the relative contribution of the symmetry energy slope to the pressure is reduced by the increasing contribution from symmetric nuclear matter. The spread in pressure at high density is therefore substantially reduced, leading to the smaller scatter at $2\rho_0$ in the right column of Fig.~\ref{fig:ds_1p4} and to an especially tight $R$-$P$ scaling relation.

We now examine systematically how the scaling relation changes with the mass of the DM candidate, retaining $\zeta_c\equiv\varepsilon_{\mathrm{DM,c}}/\varepsilon_{\mathrm{NM,c}}=1$. As shown in Table~\ref{tab:mass_scaling_full}, neither the exponent $\eta$ nor the coefficient $\bar{C}$ varies monotonically with $m_\chi$: $\eta$ develops a maximum, whereas $\bar{C}$ develops a minimum. Their combined contribution in the scaling relation $\bar{C}P^\eta$ nevertheless yields a NM radius $R_{\rm NM}$ that generally increases with the DM candidate mass $m_\chi$: as $m_\chi$ increases, the DM content decreases markedly and a larger fraction of the fixed total mass is carried by NM, thereby increasing the observable radius. This trend is illustrated by the DD2 results in Table~\ref{tab:mass_scaling_full}, where a small $m_\chi$ notably reduces the NM radius, accompanied by a larger and more extended DM component. DD2 is a density-dependent RMF model \cite{Typel2010} that is relatively stiff at high density and yields an NS maximum mass of $2.42M_\odot$ and a canonical radius $R_{1.4}=13.15\,\mathrm{km}$, consistent with constraints from massive pulsars and multimessenger observations such as GW170817.

The radius of a normal NS is governed primarily by the symmetry energy and its density dependence \cite{Lattimer2004,Lattimer2007}. Current astrophysical constraints on canonical NS radii tend to favor a softer symmetry energy \cite{Lattimer2014,Fatt2018,Li2021}. RMF models well constrained by finite-nucleus properties often predict comparatively large NS radii, and softening the symmetry energy offers a way to reduce the radii. By contrast, the parity-violating electron-scattering experiment (PREX) II favors a relatively stiff symmetry energy \cite{Adh2021}, which implies a comparatively large NS radius \cite{Reed2021}. Reconciling terrestrial experiments with astrophysical observations therefore remains challenging. Our results indicate that a GeV-scale DM candidate offers a possible mechanism for reducing the NM radius. This reduction occurs systematically across the nuclear EOS sample. We note that astrophysical determinations of neutron star radii often assume a common radius for neutron stars of different masses \cite{Lattimer2014,Guillot2014,Ozel2016}. For both NS masses considered here, the inclusion of DM reduces the NM radius and shifts the corresponding scaling parameters. These systematic departures from purely nuclear-matter results may therefore help identify the influence of DM.

As $m_\chi$ and the associated DM fraction vary, the maximum in $\eta$ and the minimum in $\bar{C}$ mark the largest departures from the no-DM baseline and are therefore the most readily distinguishable regions of the parameter scan.
At $m_\chi=600$--$550\,\mathrm{MeV}$, however, both the NM and DM radii of the $1M_\odot$ DANSs increase anomalously, accompanied by abrupt changes in the scaling exponent. For DD2, Table~\ref{tab:mass_scaling_full} gives $M_{\rm DM}/M_{\rm NM}=91.413$ and $R_{\rm NM}=15.16\,\mathrm{km}$ at $m_\chi=600\,\mathrm{MeV}$, and $M_{\rm DM}/M_{\rm NM}=15.582$ and $R_{\rm NM}=77.35\,\mathrm{km}$ at $m_\chi=550\,\mathrm{MeV}$. The corresponding values are $\eta=0.1930$ and $3.6564$, and $\bar{C}(2\rho_0)=8.118$ and $1.08\times10^{-3}$, respectively. In these configurations, DM dominates the total mass and the central NM number density falls below the crust-core transition density. The NM component is therefore no longer uniform matter near saturation density, and the scaling relation constructed from the uniform-matter pressure ceases to apply in this low density range. We note that similar extended halos formed by light DM have been reported in other DANS calculations \cite{Ivan2020,Rout2025}.

\begin{widetext}
\begin{center}
\begin{table}[tbh]
\caption{Scaling coefficient $\bar{C}(2\rho_0)$ and exponent $\eta$ for fermionic DANSs as functions of the DM particle mass. Results are given for $1M_\odot$ and $1.4M_\odot$ DANSs at $\zeta_c\equiv\varepsilon_{\mathrm{DM,c}}/\varepsilon_{\mathrm{NM,c}}=1$. The table also lists the structural quantities $R_{\rm NM}$, $R_{\rm DM}$, and $M_{\rm DM}/M_{\rm NM}$ obtained with the DD2.}
    \label{tab:mass_scaling_full}
    \fontsize{8.5pt}{9pt}\selectfont
    \begin{tabular*}{\textwidth}{@{\extracolsep{\fill}}lcccccccc@{}}
        \toprule
 & \multicolumn{4}{c}{\textbf{$1.0 M_\odot$}} & \multicolumn{4}{c}{\textbf{$1.4 M_\odot$}} \\
\cmidrule(lr){2-5} \cmidrule(lr){6-9}
\textbf{$m_\chi$} & \textbf{$R_{\rm NM}/R_{\rm DM}$} & \textbf{$M_{\rm DM}/M_{\rm NM}$} & \textbf{$\bar{C}(2\rho_0)$} & \textbf{$\eta$}
 & \textbf{$R_{\rm NM}/R_{\rm DM}$} & \textbf{$M_{\rm DM}/M_{\rm NM}$} & \textbf{$\bar{C}(2\rho_0)$} & \textbf{$\eta$} \\
\multicolumn{1}{c}{(MeV)} & (km) & & & & (km) & & & \\ \midrule
No DM & 13.07/0.00 & 0.000 & 3.375 & 0.4153 & 13.15/0.00 & 0.000 & 2.986 & 0.4534 \\
550 & 77.35/107.00 & 15.582 & 0.001 & 3.6564 & 10.73/13.01 & 0.475 & 1.833 & 0.5427 \\
600 & 15.16/54.14 & 91.413 & 8.118 & 0.1930 & 11.09/10.95 & 0.344 & 1.732 & 0.5703 \\
650 & 10.36/13.55 & 0.605 & 2.561 & 0.4275 & 11.39/9.86 & 0.267 & 1.861 & 0.5554 \\
700 & 10.80/11.17 & 0.402 & 2.600 & 0.4358 & 11.65/9.09 & 0.213 & 1.900 & 0.5558 \\
750 & 11.14/9.85 & 0.299 & 2.660 & 0.4384 & 11.88/8.47 & 0.172 & 1.964 & 0.5514 \\
800 & 11.42/8.98 & 0.232 & 2.728 & 0.4385 & 12.08/7.93 & 0.140 & 2.038 & 0.5448 \\
850 & 11.66/8.32 & 0.185 & 2.793 & 0.4379 & 12.24/7.45 & 0.115 & 2.115 & 0.5375 \\
900 & 11.87/7.77 & 0.150 & 2.859 & 0.4363 & 12.39/7.02 & 0.095 & 2.190 & 0.5301 \\
950 & 12.05/7.30 & 0.122 & 2.921 & 0.4342 & 12.51/6.63 & 0.080 & 2.271 & 0.5217 \\
1000 & 12.19/6.87 & 0.101 & 2.978 & 0.4320 & 12.61/6.28 & 0.067 & 2.344 & 0.5144 \\
1100 & 12.42/6.14 & 0.071 & 3.071 & 0.4285 & 12.77/5.65 & 0.048 & 2.472 & 0.5016 \\
1200 & 12.58/5.53 & 0.051 & 3.145 & 0.4252 & 12.86/5.11 & 0.035 & 2.586 & 0.4903 \\
1300 & 12.69/5.02 & 0.038 & 3.203 & 0.4226 & 12.93/4.65 & 0.026 & 2.673 & 0.4819 \\
1400 & 12.77/4.58 & 0.029 & 3.244 & 0.4207 & 12.98/4.25 & 0.020 & 2.741 & 0.4755 \\
1500 & 12.82/4.20 & 0.022 & 3.274 & 0.4194 & 13.02/3.90 & 0.015 & 2.791 & 0.4709 \\
1600 & 12.87/3.86 & 0.017 & 3.296 & 0.4184 & 13.04/3.59 & 0.012 & 2.830 & 0.4674 \\
1700 & 12.90/3.57 & 0.013 & 3.313 & 0.4177 & 13.07/3.33 & 0.009 & 2.860 & 0.4646 \\
1800 & 12.93/3.32 & 0.011 & 3.326 & 0.4171 & 13.08/3.09 & 0.008 & 2.884 & 0.4624 \\
1900 & 12.95/3.09 & 0.009 & 3.334 & 0.4168 & 13.09/2.88 & 0.006 & 2.903 & 0.4607 \\
2000 & 12.97/2.89 & 0.007 & 3.341 & 0.4166 & 13.10/2.69 & 0.005 & 2.919 & 0.4594 \\
\bottomrule
    \end{tabular*}
\end{table}
\end{center}
\end{widetext}

\begin{figure}[tbh]
    \centering
    \includegraphics[width=1.0\columnwidth,height=1.0\columnwidth]{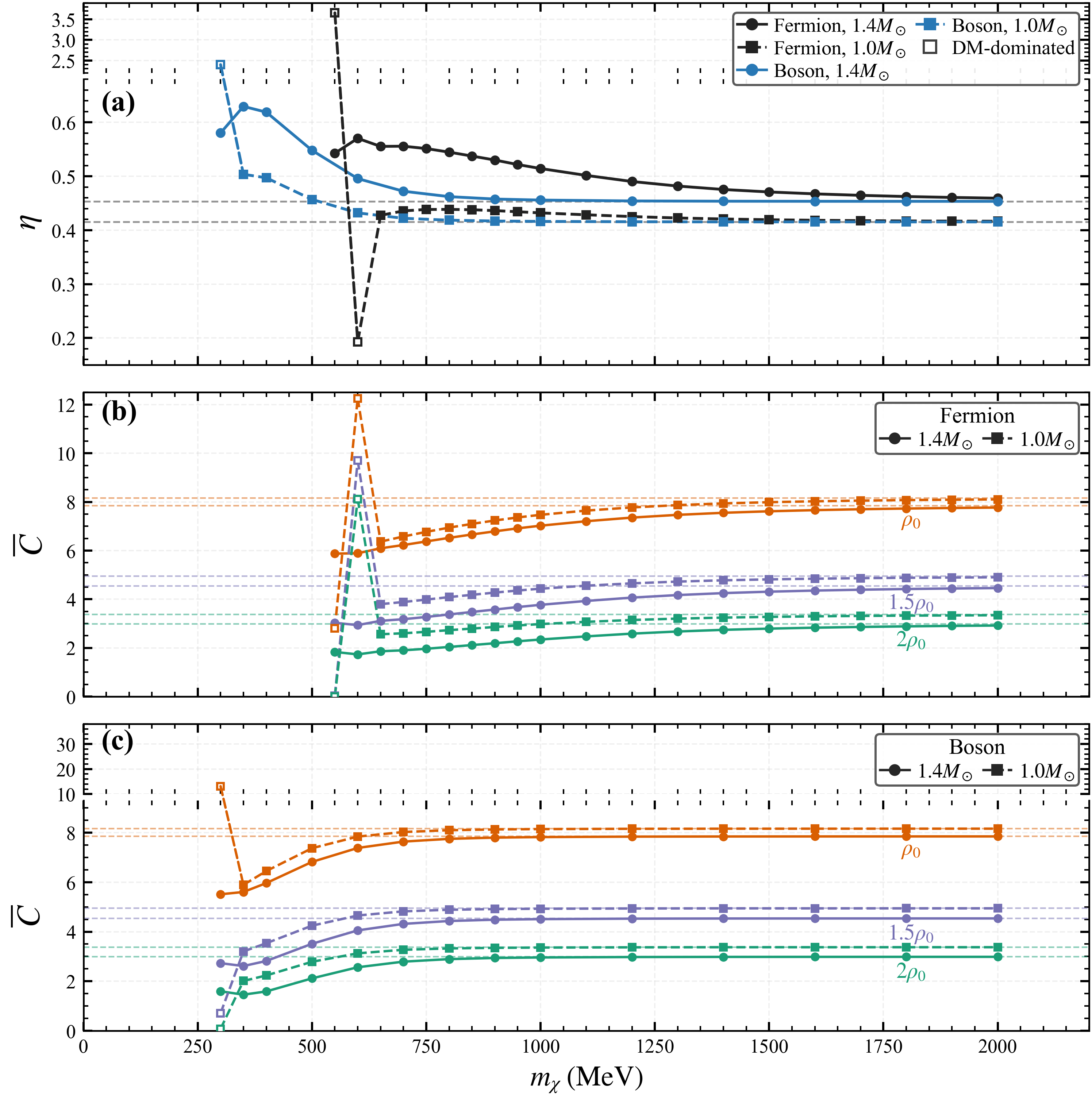}
    \caption{Scaling parameters of fermionic and bosonic DANSs as functions of the DM particle mass. (a) Scaling exponent $\eta$ for $1.4M_\odot$ and $1.0M_\odot$ DANSs. Open squares denote transitions to DM-dominated configurations, and the horizontal dashed lines indicate the no-DM baselines. The two anomalous low-mass exponents are shown above the break in the vertical axis. (b), (c) Scaling coefficients $\bar{C}$ at $\rho_0$, $1.5\rho_0$, and $2\rho_0$ for fermionic and bosonic DM, respectively. The horizontal dashed lines indicate the no-DM baselines.}
    \label{fig:mass_scaling_parameters}
\end{figure}

We next examine the particle-mass dependence of the bosonic DM (BDM) model. Figure~\ref{fig:mass_scaling_parameters}(a) compares the scaling exponents for fermionic and bosonic DM, while Figs.~\ref{fig:mass_scaling_parameters}(b) and \ref{fig:mass_scaling_parameters}(c) show the corresponding coefficients $\bar{C}$ for the fermionic and bosonic DM, respectively.   We scan the mass of the bosonic DM candidate from $300$ to $2000\,\mathrm{MeV}$. For the $1.4M_\odot$ bosonic DANSs, the exponent varies nonmonotonically with mass and develops a maximum, while the coefficient develops a minimum at lower $m_\chi$. For the $1.0M_\odot$ sequence, by contrast, $\eta$ decreases monotonically and $\bar{C}$ increases monotonically with $m_\chi$ along the regular-radius branch. At $m_\chi=300\,\mathrm{MeV}$, the $1.0M_\odot$ bosonic DANS becomes DM dominated and its central NM density falls below the crust-core transition density. With the DD2, this configuration has $R_{\rm NM}=114.15\,\mathrm{km}$, $R_{\rm DM}=132.34\,\mathrm{km}$, and $M_{\rm DM}/M_{\rm NM}=4.9968$, with $\eta=2.4002$ and $\bar{C}(2\rho_0)=0.0780$.
Thus, most of the total mass is carried by DM, while the nonuniform NM component extends beyond $100\,\mathrm{km}$. For both $1.0M_\odot$ and $1.4M_\odot$ bosonic DANSs, appreciable departures of the scaling exponent and coefficient from the no-DM values occur for $m_\chi<700\,\mathrm{MeV}$ and grow as the DM candidate mass decreases. This explicit mass dependence provides a potential signature of light bosonic DM.

\subsection{Effects of the DM Content on the DANS Radius and Scaling Relation}
The preceding analysis varied the DM particle mass at fixed $\zeta_c=\varepsilon_{\mathrm{DM,c}}/\varepsilon_{\mathrm{NM,c}}=1$. We now fix the particle mass and vary the DM content of the DANS by changing the central dark-to-NM energy-density ratio $\zeta_c$. The DM candidate mass is fixed at $m_\chi=1\,\mathrm{GeV}$ and the self-interactions are set to zero. Table~\ref{tab:ratio_scaling_full} presents the fermionic DANS results for $0\leq\zeta_c\leq3.0$. The values of $\eta$ and $\bar{C}(2\rho_0)$ are fitted using all 21 EOSs, while the structural quantities $R_{\rm NM}$, $R_{\rm DM}$, and $M_{\rm DM}/M_{\rm NM}$ are obtained with the DD2.

Increasing $\zeta_c$ increases the DM content of the DANS. For DD2, $M_{\rm DM}/M_{\rm NM}$ increases continuously with $\zeta_c$. At fixed total mass, the NM mass consequently decreases, and $R_{\rm NM}$ decreases monotonically. For the $1.0M_\odot$ DANSs, $\eta$ increases from the no-DM value of $0.4153$ to $0.4657$ at $\zeta_c=2.5$, before decreasing slightly to $0.4565$ at $\zeta_c=3.0$; $\bar{C}(2\rho_0)$ decreases from $3.375$ to $2.392$ and then remains nearly constant. For the $1.4M_\odot$ configurations, $\eta$ increases from $0.4534$ to $0.5898$ over the scanned range, whereas $\bar{C}(2\rho_0)$ decreases continuously to $1.674$. The different changing  rates  of the scaling parameters are exhibited in two target DANS with different masses, indicating that the DM matter effect on the scaling relation not only depends on the DM content but also on the total stellar mass.

\begin{widetext}
\begin{center}
\begin{table}[tbh]
\centering
\caption{Scaling exponent $\eta$ and coefficient $\bar{C}$ of fermionic DANSs as functions of the central dark-to-NM energy-density ratio $\zeta_c$. Results are given for $1M_\odot$ and $1.4M_\odot$ DANSs with $m_\chi=1\,\mathrm{GeV}$ and no DM self-interaction. The coefficient $\bar{C}$ is fitted at $2\rho_0$. The table also lists the structural properties $R_{\rm NM}$, $R_{\rm DM}$, and $M_{\rm DM}/M_{\rm NM}$ obtained with the DD2 model.}
\label{tab:ratio_scaling_full}
\fontsize{8.5pt}{9pt}\selectfont
\begin{tabular*}{\textwidth}{@{\extracolsep{\fill}}lcccccccc@{}}
\toprule
 & \multicolumn{4}{c}{\textbf{$1.0 M_\odot$}} & \multicolumn{4}{c}{\textbf{$1.4 M_\odot$}} \\
\cmidrule(lr){2-5} \cmidrule(lr){6-9}
\textbf{$\zeta_c$} & \textbf{$R_{\rm NM}/R_{\rm DM}$} & \textbf{$M_{\rm DM}/M_{\rm NM}$} & \textbf{$\bar{C}$} & \textbf{$\eta$}
 & \textbf{$R_{\rm NM}/R_{\rm DM}$} & \textbf{$M_{\rm DM}/M_{\rm NM}$} & \textbf{$\bar{C}$} & \textbf{$\eta$} \\
 & (km) & & & & (km) & & & \\ \midrule
0.0 (No DM) & 13.07/0.00 & 0.000 & 3.375 & 0.4153 & 13.15/0.00 & 0.000 & 2.986 & 0.4534 \\
0.1 & 13.04/4.19 & 0.002 & 3.365 & 0.4157 & 13.14/3.82 & 0.001 & 2.970 & 0.4548 \\
0.2 & 12.98/5.06 & 0.007 & 3.341 & 0.4166 & 13.10/4.63 & 0.005 & 2.930 & 0.4581 \\
0.3 & 12.90/5.59 & 0.015 & 3.311 & 0.4176 & 13.06/5.12 & 0.010 & 2.872 & 0.4632 \\
0.4 & 12.81/5.95 & 0.025 & 3.273 & 0.4190 & 13.00/5.46 & 0.017 & 2.801 & 0.4695 \\
0.5 & 12.72/6.22 & 0.036 & 3.230 & 0.4206 & 12.94/5.71 & 0.024 & 2.725 & 0.4764 \\
0.6 & 12.62/6.43 & 0.048 & 3.184 & 0.4225 & 12.88/5.90 & 0.032 & 2.645 & 0.4839 \\
0.7 & 12.51/6.58 & 0.060 & 3.131 & 0.4249 & 12.82/6.04 & 0.040 & 2.564 & 0.4919 \\
0.8 & 12.41/6.70 & 0.074 & 3.077 & 0.4275 & 12.75/6.14 & 0.049 & 2.483 & 0.5000 \\
0.9 & 12.30/6.80 & 0.087 & 3.026 & 0.4299 & 12.69/6.22 & 0.058 & 2.406 & 0.5081 \\
1.0 & 12.19/6.87 & 0.101 & 2.978 & 0.4320 & 12.61/6.28 & 0.067 & 2.344 & 0.5144 \\
1.2 & 11.98/6.97 & 0.130 & 2.886 & 0.4363 & 12.46/6.35 & 0.085 & 2.222 & 0.5274 \\
1.4 & 11.78/7.02 & 0.159 & 2.820 & 0.4381 & 12.31/6.38 & 0.103 & 2.123 & 0.5381 \\
1.6 & 11.60/7.03 & 0.188 & 2.746 & 0.4414 & 12.18/6.39 & 0.120 & 2.030 & 0.5486 \\
1.8 & 11.45/7.02 & 0.216 & 2.648 & 0.4481 & 12.06/6.37 & 0.137 & 1.949 & 0.5582 \\
2.0 & 11.32/6.99 & 0.244 & 2.551 & 0.4554 & 11.96/6.34 & 0.153 & 1.880 & 0.5663 \\
2.5 & 11.01/6.86 & 0.310 & 2.392 & 0.4657 & 11.71/6.22 & 0.191 & 1.739 & 0.5839 \\
3.0 & 10.66/6.74 & 0.371 & 2.395 & 0.4565 & 11.48/6.07 & 0.223 & 1.674 & 0.5898 \\
\bottomrule
\end{tabular*}
\end{table}
\end{center}
\end{widetext}

\clearpage
\begin{figure}[tbh]
\centering
    \includegraphics[width=\columnwidth]{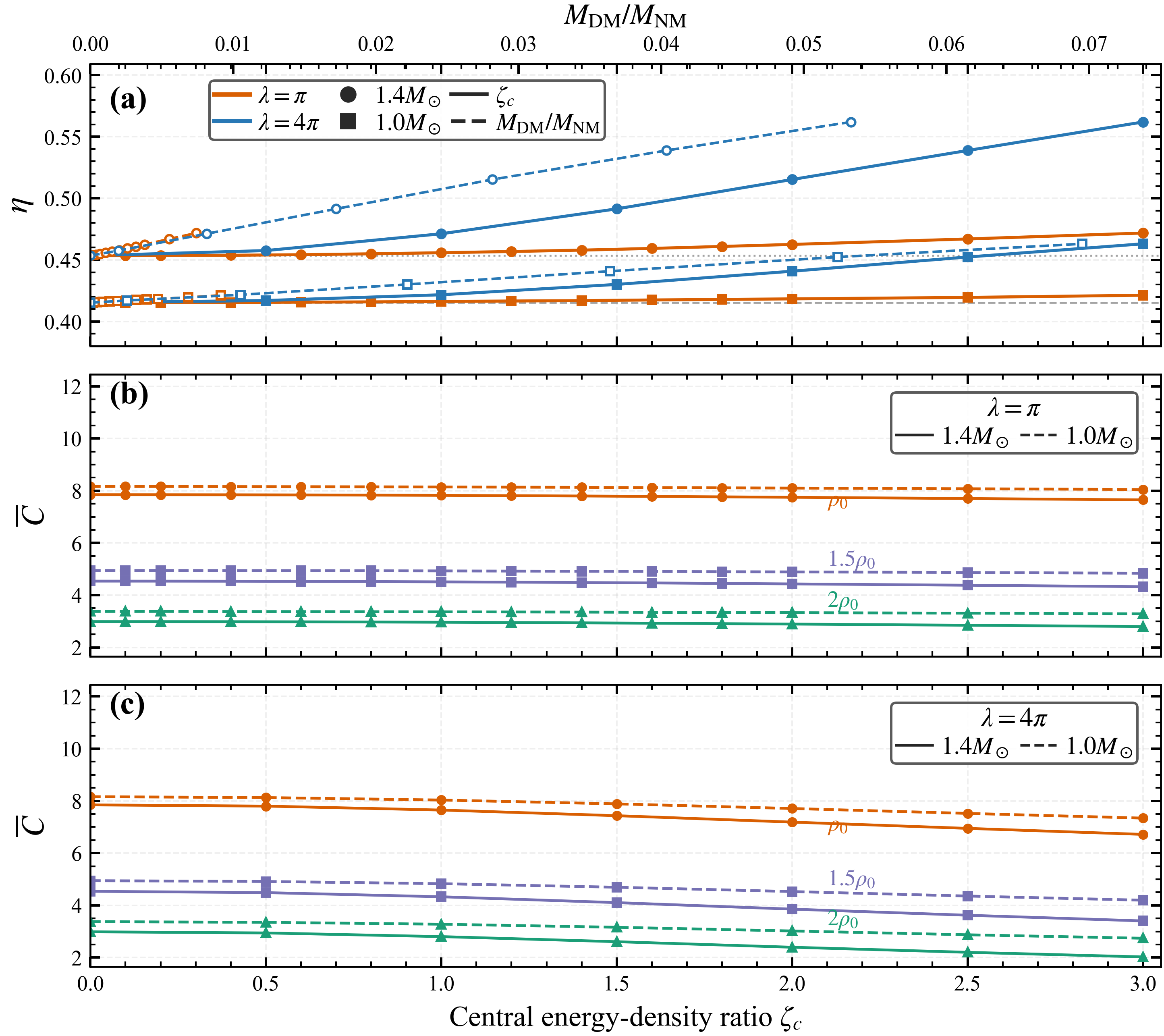}
    \caption{Scaling parameters of bosonic DANSs as functions of the central dark-to-NM energy-density ratio $\zeta_c$. (a) Scaling exponent $\eta$. Orange and blue curves correspond to $\lambda=\pi$ and $4\pi$, respectively, and circles and squares denote the $1.4M_\odot$ and $1.0M_\odot$ configurations. Dashed curves refer to the upper axis and show the dark-to-NM mass ratio $M_{\rm DM}/M_{\rm NM}$ with the DD2. (b), (c) Scaling coefficients $\bar{C}(\rho_k)$ for $\lambda=\pi$ and $4\pi$, respectively. Solid and dashed curves correspond respectively to the $1.4M_\odot$ and $1.0M_\odot$ configurations, while circles, squares, and triangles denote $\rho_0$, $1.5\rho_0$, and $2\rho_0$, respectively. }
\label{fig:eta_C_RNM_compare}
\end{figure}

Figure~\ref{fig:eta_C_RNM_compare} shows the variation of the BDM scaling parameters with $\zeta_c$. In contrast to the fermionic results in Table~\ref{tab:ratio_scaling_full}, the BDM scaling exponent increases monotonically with $\zeta_c$. This difference is primarily associated with the repulsive self-interaction included in the BDM model. Figure~\ref{fig:eta_C_RNM_compare}(a) also shows that the scaling parameters depend differently on $M_{\rm DM}/M_{\rm NM}$ and on $\zeta_c$. To examine the effect of the repulsive bosonic self-interaction, Fig.~\ref{fig:eta_C_RNM_compare} compares $\lambda=\pi$ and $4\pi$ at fixed $m_\chi=1\,\mathrm{GeV}$. Note that the larger value of $\lambda$ represents stronger repulsion. For $\lambda=\pi$, the changes in $\eta$ and the three coefficients $\bar{C}(\rho_k)$ are relatively small. When the coupling is increased to $\lambda=4\pi$, $\eta$ rises more rapidly with the DM content, while all three coefficients decrease. In fact, the more explicit change with larger $\lambda$ is associated with the enhanced change in the mass ratio $M_{\rm DM}/M_{\rm NM}$. At $m_\chi=1\,\mathrm{GeV}$, bosonic DM tends to form a compact dark core and shifts the mass--radius relation toward smaller radii; stronger repulsive self-interactions partially counteract this structural change \cite{Giangrandi2023,Kark2022}.

\subsection{Effects of Fermionic DM Self-Interactions on the Scaling Parameters}

Having examined the effect of the bosonic self-interaction above, we now consider how fermionic DM self-interactions modify the $R$-$P$ scaling relation. The fermionic Lagrangian in Eq.~\eqref{eq:FermionL} contains scalar and vector mediators, which generate attractive and repulsive interactions, respectively. Such scalar-vector mean-field models have been widely used to study the structure of self-interacting dark stars and DANSs \cite{Narain2006,Xiang2014,Moira2019,Miao2022,Das2020_DANS_GW}.  We set both mediator masses to $1\,\mathrm{GeV}$ and vary the corresponding coupling strengths $C_s$ and $C_v$, to determine how attraction and repulsion modify the scaling relation.
\begin{figure}[tbh]
    \centering
    \includegraphics[width=\columnwidth]{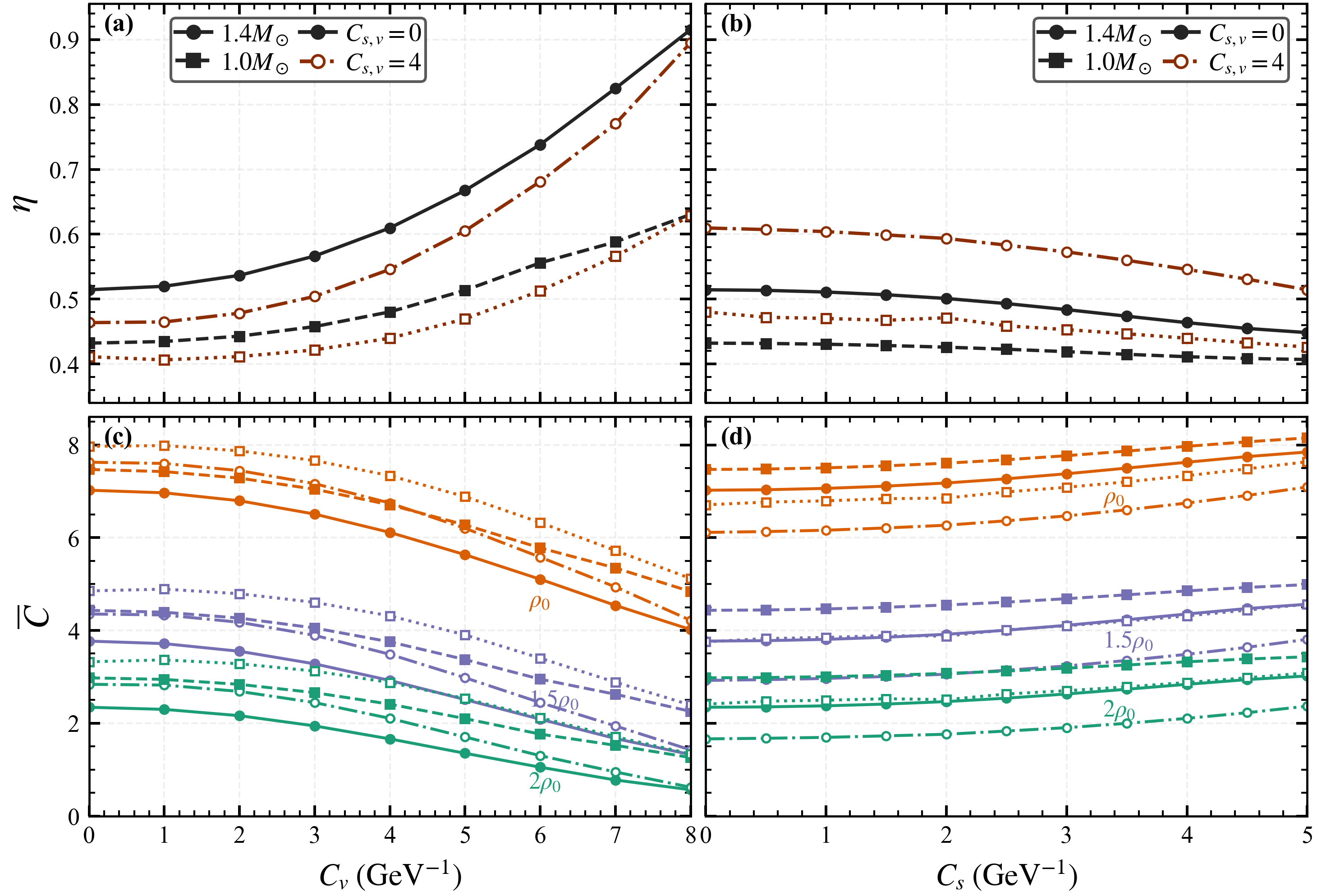}
    \caption{Scaling exponent and coefficients of fermionic DANSs as functions of the effective DM self-interaction strengths $C_s$ and $C_v$. (a) Scaling exponent $\eta$ as $C_v$  varies at fixed $C_s=0$ or $4\,\mathrm{GeV}^{-1}$. (b) Scaling exponent $\eta$ as $C_s$ varies at fixed $C_v=0$ or $4\,\mathrm{GeV}^{-1}$. (c), (d) Corresponding scaling coefficients $\bar{C}(\rho_k)$ at $\rho_0$, $1.5\rho_0$, and $2\rho_0$. Black solid or dashed curves denote results with only one interaction, while brown curves with open symbols denote results on which both interactions are present.}\label{fig:DM_interaction_scaling}
\end{figure}

Figure~\ref{fig:DM_interaction_scaling} shows the effects of the repulsive strength $C_v$ and attractive strength $C_s$ on the exponent $\eta$ and the three coefficients $\bar{C}(\rho_k)$. The curves cover three cases: scalar attraction alone, vector repulsion alone, and both interactions included simultaneously. For the latter case, we vary $C_v$ from $0$ to $8\,\mathrm{GeV}^{-1}$ at fixed $C_s=4\,\mathrm{GeV}^{-1}$, or vary $C_s$ from $0$ to $5\,\mathrm{GeV}^{-1}$ at fixed $C_v=4\,\mathrm{GeV}^{-1}$.

At fixed $C_s=0$, increasing $C_v$ raises $\eta$ overall and lowers $\bar{C}(\rho_k)$ at all three densities. When $C_s$ is fixed instead at $4\,\mathrm{GeV}^{-1}$, the $\eta$ curve shifts downward and the $\bar{C}(\rho_k)$ curves shift upward, showing that scalar attraction partially cancels  the effect of vector repulsion.

With attractive self-interactions alone, the DM distribution contracts toward the stellar center. As $C_s$ increases, the $\eta$ curve shifts downward and the $\bar{C}(\rho_k)$ curves shift upward, with relatively gradual variations. At fixed $C_v=4\,\mathrm{GeV}^{-1}$, the $\eta$ curve lies above the corresponding $C_v=0$ result. Overall, scalar attraction and vector repulsion shift the scaling exponent in opposite directions.

In the above analysis, we have scanned the particle mass, central energy-density ratio, and self-interaction strength. For a fixed nuclear EOS sample and DANS mass, the results show that  the parameters $m_\chi$, $\zeta_c$, $C_s$, and $C_v$ all shift both the exponent and coefficient of the $R$-$P$ relation. Studying these shifts for different DM scenarios connects macroscopic NS observables with the underlying properties of both nuclear matter and DM. At fixed DANS mass, DM primarily modifies the radius. While the tidal deformability is itself strongly radius dependent, more precise measurements of DANS radii and tidal deformabilities could therefore help constrain both the nuclear EOS and DM properties \cite{Kumar2024,Giangrandi2023}.

\section{Summary}

Within a two-fluid TOV framework, we have studied how fermionic and bosonic DM coupled to NM only through gravity affect the matter distributions, radii, and nuclear  pressure--observable radius scaling relations of DANSs. Following the $R$-$P$ relation established in normal NSs, we extended the Buchdahl solution to a two-fluid system and derived corresponding analytic corrections to the scaling parameters. Numerically, we established a broadly representative sample of 21 nuclear EOSs and optimized the fit to extract the scaling exponent $\eta$ and coefficient $\bar{C}$ for $1.0M_\odot$ and $1.4M_\odot$ DANSs.

We  have investigated the DM effects in DANSs systematically by varying the mass of the GeV- and sub-GeV-scale DM candidates, the central dark-to-NM energy-density ratios, and the DM self-interactions.  This systematic study not only helps evaluate how dark matter affects the astrophysical constraints on the nuclear EOS, but also allows distinguishing, through various DM scenarios, generic versus specific differences relative to normal NSs, thereby aiding the identification of the DM signals.

As the DM candidate mass increases, the dark component changes from an extended distribution to a centrally concentrated one. For both fermionic and bosonic DANSs, $\eta$ and $\bar{C}$ generally vary nonmonotonically. Over a broad mass range of DM candidates, adding DM reduces the NM radius and systematically shifts the scaling parameters. The scaling parameters of the $1.4M_\odot$ bosonic DANSs are most sensitive to the candidate mass in the sub-GeV range, whereas the fermionic-DM effects persist up to $m_\chi\simeq1500\,\mathrm{MeV}$.
 
At the low candidate mass, both types of DM can produce DM-dominated, radially extended configurations in which the central NM is at the nonuniform low-density regime. DM-dominated configurations emerge at different low-mass scales: approximately $550$--$600\,\mathrm{MeV}$ for fermionic DM and $300\,\mathrm{MeV}$ for bosonic DM in the representative $1.0M_\odot$ DANS. The scaling relation based on the pressure of uniform nuclear matter then becomes invalid, as manifested by abrupt changes in its fitted parameters. In this case, the observable NM radius also acquires a large rise, which may serve as a signal for such specific DM candidates.

We also examined the dependence on the central energy-density ratio $\zeta_c$ and the corresponding DM content of the star. Given the total mass of DANSs, the DM mass fraction increases with $\zeta_c$, and $R_{\rm NM}$ decreases monotonically. The scaling exponent generally increases and the coefficient decreases, which may develop meaningful differences from those of normal NSs, providing potential DM signals due to the rich DM embedment in DANSs.

The DM self-interaction is another factor to affect the DM effect on stellar properties.  Repulsive interactions stiffen the DM EOS and enhance the DM content in DANSs, which increases $\eta$ and decrease $\bar{C}$. Note that the DM repulsion may smear out the difference in NM radii associated with various nuclear symmetry energies. The insensitivity to the content of nuclear EOSs  may actually bring the opportunity to sharpening the DM signals. In comparison, the DM attractive interaction plays the role in the opposite direction.

In summary, our results show how various DM scenarios modify the stellar radii and the scaling  relation between the observable NM radius and nuclear pressure.  These systematic shifts in  $\eta$ and $\bar{C}$ connect the DM properties, the observable star radius, and the nuclear EOS, offering a new perspective for future studies that combine NS observations to investigate DM properties and constrain the nuclear EOS.
\clearpage
\appendix
\section{\texorpdfstring{Two-Fluid Extension of the Buchdahl Solution and the $R$-$P$ Scaling Relation}{Two-Fluid Extension of the Buchdahl Solution and the R-P Scaling Relation}}
\label{app:buchdahl_scaling}

This Appendix derives the $R$-$P$ scaling exponent of a two-fluid system within the Buchdahl analytic model. In a static, spherically symmetric spacetime, the NM and DM energy--momentum tensors add to give the total pressure and energy density,
\begin{equation}
P_t=P_{\rm NM}+P_{\rm DM},
\qquad
\varepsilon_t=\varepsilon_{\rm NM}+\varepsilon_{\rm DM}.
\label{eq:app_total_fluid}
\end{equation}
In the following, the two static perfect fluids are treated as an effective isotropic fluid whose analytic structure is described by the Buchdahl relation. We use units with $G=c=1$.

\subsection{Buchdahl Relation for the Total Fluid}

Following the metric-signature convention $(+,-,-,-)$ of Buchdahl, the line element is written as
\begin{equation}
ds^2=-e^{\lambda(r)}dr^2-r^2e^{\mu(r)}d\Omega^2+e^{\nu(r)}dt^2,
\label{eq:app_buch_metric}
\end{equation}
where $r$ is a radial coordinate and the corresponding areal radius is
\begin{equation}
r_{\rm area}=r e^{\mu(r)/2}.
\label{eq:app_area_radius}
\end{equation}
After imposing regularity at the center and matching to the exterior Schwarzschild solution, the Buchdahl solution gives \cite{Buchdahl1967,Lattimer2001}
\begin{equation}
\varepsilon_t=12\sqrt{p_{*,t}P_t}-5P_t,
\label{eq:app_dm_buch_eos}
\end{equation}
and the total-fluid boundary expressed in terms of the areal radius,
\begin{equation}
R_t=(1-\beta_t)
\left[
\frac{\pi}{288p_{*,t}(1-2\beta_t)}
\right]^{1/2},
\qquad
\beta_t=\frac{M}{R_t}.
\label{eq:app_buch_Rt}
\end{equation}
where $M$ is the total gravitational mass and $p_{*,t}$ is a scale parameter with dimensions of pressure. These two relations form the starting point for the derivatives below.

\subsection{Total-Fluid Boundary and Normal-Matter Radius}

The outer boundary is $R_t=\max(R_{\rm NM},R_{\rm DM})$, whereas the electromagnetically observable boundary is $R_{\rm NM}$. To express the total-fluid relation in terms of the NM radius, we define
\begin{equation}
R_{\rm NM}=\gamma R_t,
\qquad
\beta=\frac{M}{R_{\rm NM}},
\qquad
\beta_t=\gamma\beta,
\label{eq:app_gamma_def}
\end{equation}
where $0<\gamma\leq1$. An extended DM halo corresponds to $\gamma<1$, whereas $\gamma=1$ when the DM component lies entirely within the NM component.

\subsection{Scaling Exponent}

We consider the local pressure response at fixed $M$, reference total energy density $\varepsilon_t$, and boundary ratio $\gamma$. Equation~\eqref{eq:app_dm_buch_eos} gives
\begin{equation}
\left.
\frac{d\ln p_{*,t}}{d\ln P_t}
\right|_{\varepsilon_t}
=-\left(
1-\frac{5}{6}\sqrt{\frac{P_t}{p_{*,t}}}
\right).
\label{eq:app_pressure_scale_derivative}
\end{equation}
The two components are described by separate EOSs. For this local response, we hold $P_{\rm DM}$ fixed. It then follows directly from $P_t=P_{\rm NM}+P_{\rm DM}$ that $dP_t=dP_{\rm NM}$ and hence
\begin{equation}
\frac{d\ln P_t}{d\ln P_{\rm NM}}
=\frac{P_{\rm NM}}{P_t}.
\label{eq:app_pressure_response}
\end{equation}
Applying the chain rule yields
\begin{equation}
\left.
\frac{d\ln p_{*,t}}{d\ln P_{\rm NM}}
\right|_{\varepsilon_t}
=-\left(
1-\frac{5}{6}\sqrt{\frac{P_t}{p_{*,t}}}
\right)\frac{P_{\rm NM}}{P_t}.
\label{eq:app_pressure_nm_derivative}
\end{equation}

Under the fixed conditions specified above, we define the scaling exponent as
\begin{equation}
\eta\equiv
\left.
\frac{d\ln R_{\rm NM}}{d\ln P_{\rm NM}}
\right|_{\varepsilon_t,M}.
\label{eq:app_eta_definition}
\end{equation}
Since $\beta_t=M/R_t=\gamma M/R_{\rm NM}$, fixing $M$ and $\gamma$ gives
\begin{equation}
\frac{d\beta_t}{d\ln P_{\rm NM}}=-\beta_t\eta.
\label{eq:app_compactness_derivative}
\end{equation}
Taking the logarithm of Eq.~\eqref{eq:app_buch_Rt} and differentiating with respect to $P_{\rm NM}$ gives
\begin{equation}
\eta=
-\frac{1}{2}\frac{d\ln p_{*,t}}{d\ln P_{\rm NM}}
+\frac{\beta_t}{(1-\beta_t)(1-2\beta_t)}
\frac{d\beta_t}{d\ln P_{\rm NM}}.
\label{eq:app_eta_intermediate}
\end{equation}
Substituting Eqs.~\eqref{eq:app_pressure_nm_derivative} and \eqref{eq:app_compactness_derivative} and rearranging, we obtain
\begin{equation}
\eta=
\frac{1}{2}\frac{P_{\rm NM}}{P_t}
\left(
1-\frac{5}{6}\sqrt{\frac{P_t}{p_{*,t}}}
\right)
\frac{(1-\beta_t)(1-2\beta_t)}
{1-3\beta_t+3\beta_t^2}.
\label{eq:app_eta_general}
\end{equation}
Finally, substituting $P_t=P_{\rm NM}+P_{\rm DM}$ and $\beta_t=\gamma\beta$ gives the two-fluid scaling exponent used in the main text,
\begin{equation}
\eta=
\frac{1}{2}
\frac{P_{\rm NM}}{P_{\rm NM}+P_{\rm DM}}
\left(
1-\frac{5}{6}\sqrt{\frac{P_t}{p_{*,t}}}
\right)
\frac{(1-\gamma\beta)(1-2\gamma\beta)}
{1-3\gamma\beta+3\gamma^2\beta^2}.
\label{eq:app_dm_scaling_used}
\end{equation}

Equation~\eqref{eq:app_dm_scaling_used} shows that DM modifies the scaling exponent through the NM fraction of the total pressure, the total-fluid pressure scale $p_{*,t}$, and the compactness at the total boundary, $\beta_t=\gamma\beta$. In the absence of DM, $P_t\rightarrow P_{\rm NM}$ and $\gamma\rightarrow1$, and the expression reduces to the single-fluid result. This local analytic relation identifies the terms responsible for the modification, whereas the scaling parameters in the main text are obtained by fitting the EOS sample calculated with the full two-fluid TOV equations.

\clearpage
\bibliography{myrefs}

@article{Lattimer2007,
  author = {Lattimer, J. M. and Prakash, M.},
  title = {Neutron star observations: Prognosis for equation of state constraints},
  journal = {Physics Reports},
  volume = {442},
  pages = {109--165},
  year = {2007}
}

@article{Abbott2018,
  title = {GW170817: Measurements of Neutron Star Radii and Equation of State},
  author = {Abbott, B. P. and others},
  collaboration = {LIGO Scientific Collaboration and Virgo Collaboration},
  journal = {Phys. Rev. Lett.},
  volume = {121},
  issue = {16},
  pages = {161101},
  year = {2018},
  publisher = {American Physical Society}
}

@article{Oppenheimer1939,
  author = {Oppenheimer, J. R. and Volkoff, G. M.},
  title = {On Massive Neutron Cores},
  journal = {Physical Review},
  volume = {55},
  number = {4},
  pages = {374--381},
  year = {1939},
  doi = {10.1103/PhysRev.55.374},
  publisher = {American Physical Society}
}

@article{Akmal1998,
  title = {Equation of state of nucleon matter and neutron star structure},
  author = {Akmal, A. and Pandharipande, V. R. and Ravenhall, D. G.},
  journal = {Physical Review C},
  volume = {58},
  number = {3},
  pages = {1804},
  year = {1998},
  doi = {10.1103/PhysRevC.58.1804},
  publisher = {American Physical Society}
}

@article{Oertel2017,
  author = {Oertel, M. and Hempel, M. and Kl{\"a}hn, T. and Typel, S.},
  title = {Equations of state for supernovae and compact stars},
  journal = {Reviews of Modern Physics},
  volume = {89},
  pages = {015007},
  year = {2017}
}

@article{Glendenning1992q,
  author  = {Glendenning, Norman K. and Weber, Fridolin},
  title   = {Nuclear solid crust on rotating strange quark stars},
  journal = {The Astrophysical Journal},
  volume  = {400},
  pages   = {647--658},
  year    = {1992},
  doi     = {10.1086/172026}
}

@article{Alford2008,
  title={Color superconductivity in dense quark matter},
  author={Alford, Mark G and Schmitt, Andreas and Rajagopal, Krishna and Sch{\"a}fer, Thomas},
  journal={Reviews of Modern Physics},
  volume={80},
  number={4},
  pages={1455},
  year={2008},
  publisher={APS}
}

@article{Zhang2024,
  author  = {Zhang, Xiao-Li and Huang, Yong-Feng and Zou, Ze-Cheng},
  title   = {Recent progresses in strange quark stars},
  journal = {Frontiers in Astronomy and Space Sciences},
  volume  = {11},
  pages   = {1409463},
  year    = {2024},
  doi     = {10.3389/fspas.2024.1409463},
}

@article{Sandin2009,
  author = {Sandin, F. and Ciarcelluti, P.},
  title = {Effects of mirror dark matter on neutron stars},
  journal = {Astroparticle Physics},
  volume = {32},
  pages = {278--284},
  year = {2009}
}

@article{Leung2011,
  author = {Leung, S. C. and Chu, M. C. and Lin, L. M.},
  title = {Dark-matter admixed neutron stars},
  journal = {Physical Review D},
  volume = {84},
  number = {10},
  pages = {107301},
  year = {2011},
  doi = {10.1103/PhysRevD.84.107301}
}

@article{Xiang2014,
  author = {Xiang, Qian-Fei and Jiang, Wei-Zhou and Zhang, Dong-Rui and Yang, Rong-Yao},
  title = {Effects of fermionic dark matter on properties of neutron stars},
  journal = {Physical Review C},
  volume = {89},
  pages = {025803},
  year = {2014}
}

@article{Day1967,
  author  = {Day, B. D.},
  title   = {Elements of the Brueckner-Goldstone Theory of Nuclear Matter},
  journal = {Rev. Mod. Phys.},
  volume  = {39},
  pages   = {719--744},
  year    = {1967},
  doi     = {10.1103/RevModPhys.39.719}
}

@article{Br1990,
  title = {Relativistic nuclear structure. I. Nuclear matter},
  author = {Brockmann, R. and Machleidt, R.},
  journal = {Phys. Rev. C},
  volume = {42},
  issue = {5},
  pages = {1965--1980},
  numpages = {0},
  year = {1990},
  month = {Nov},
  doi = {10.1103/PhysRevC.42.1965},
  url = {https://link.aps.org/doi/10.1103/PhysRevC.42.1965}
}

@article{Song1998,
  author  = {Song, H. Q. and Baldo, M. and Giansiracusa, G. and Lombardo, U.},
  title   = {Bethe-Brueckner-Goldstone Expansion in Nuclear Matter},
  journal = {Phys. Rev. Lett.},
  volume  = {81},
  number  = {8},
  pages   = {1584--1587},
  year    = {1998},
  doi     = {10.1103/PhysRevLett.81.1584}
}

@article{Wiringa1988,
  author = {Wiringa, R. B. and Fiks, V. and Fabrocini, A.},
  title = {Equation of state for dense nucleon matter},
  journal = {Physical Review C},
  volume = {38},
  number = {2},
  pages = {1010--1037},
  year = {1988},
  doi = {10.1103/PhysRevC.38.1010}
}

@ARTICLE{Di2003,
       author = {{Dieperink}, A.~E.~L. and {Dewulf}, Y. and {van Neck}, D. and {Waroquier}, M. and {Rodin}, V.},
        title = "{Nuclear symmetry energy and the neutron skin in neutron-rich nuclei}",
      journal = {Physical Review C},
     year = 2003,
        month = dec,
       volume = {68},
       number = {6},
          eid = {064307},
        pages = {064307},
          doi = {10.1103/PhysRevC.68.064307}
}

@ARTICLE{Fu1996,
       author = {{Furnstahl}, R.~J. and {Tang}, H.~B. and {Serot}, B.~D.},
        title = "{Vacuum contributions in a chiral effective Lagrangian for nuclei}",
      journal = {Physical Review C},
         year = 1995,
        month = sep,
       volume = {52},
       number = {3},
        pages = {1368-1379},
          doi = {10.1103/PhysRevC.52.1368}
}

@ARTICLE{Fi2006,
       author = {{Finelli}, P. and {Kaiser}, N. and {Vretenar}, D. and {Weise}, W.},
        title = "{Relativistic nuclear energy density functional constrained by low-energy QCD}",
      journal = {Nuclear Physics A},
         year = 2006,
        month = may,
       volume = {770},
       number = {1-2},
        pages = {1-31},
          doi = {10.1016/j.nuclphysa.2006.02.007}
}

@article{Hebeler2013,
  title={Equation of state and neutron star properties constrained by nuclear physics and observation},
  author={Hebeler, K. and Lattimer, J. M. and Pethick, C. J. and Schwenk, A.},
  journal={The Astrophysical Journal},
  volume={773},
  number={1},
  pages={11},
  year={2013},
  publisher={IOP Publishing}
}

@article{Walecka1974,
  title={A theory of highly condensed matter},
  author={Walecka, J. D.},
  journal={Annals of Physics},
  volume={83},
  number={2},
  pages={491--529},
  year={1974},
  publisher={Elsevier}
}

@article{Dutra2014,
  title={Relativistic mean-field hadronic models under nuclear matter constraints},
  author={Dutra, M. and Louren{\c{c}}o, O. and Avancini, S. S. and Carlson, B. V. and Delfino, A. and Menezes, D. P. and Provid{\^e}ncia, C. and Typel, S. and Xia, C. R.},
  journal={Physical Review C},
  volume={90},
  number={5},
  pages={055203},
  year={2014},
  publisher={APS}
}

@ARTICLE{Li2008,
       author = {{Li}, Bao-An and {Chen}, Lie-Wen and {Ko}, Che Ming},
        title = "{Recent progress and new challenges in isospin physics with heavy-ion reactions}",
      journal = {Physics Reports},
         year = 2008,
        month = aug,
       volume = {464},
       number = {4-6},
        pages = {113-281},
          doi = {10.1016/j.physrep.2008.04.005}
}

@article{Lattimer2012,
  title={The nuclear equation of state and neutron star masses},
  author={Lattimer, James M},
  journal={Annual Review of Nuclear and Particle Science},
  volume={62},
  pages={485--515},
  year={2012},
  publisher={Annual Reviews},
  url={https://www.annualreviews.org/doi/abs/10.1146/annurev-nucl-102711-095018}
}

@ARTICLE{Lattimer2004,
       author = {{Lattimer}, J.~M. and {Prakash}, M.},
        title = "{The physics of neutron stars}",
      journal = {Science},
         year = 2004,
        month = apr,
       volume = {304},
       number = {5670},
        pages = {536-542},
          doi = {10.1126/science.1090720}
}

@article{Glendenning1992,
  title={First-order phase transitions with more than one conserved charge: Consequences for neutron stars},
  author={Glendenning, Norman K},
  journal={Physical Review D},
  volume={46},
  number={4},
  pages={1274},
  year={1992},
  publisher={APS}
}

@ARTICLE{Jiang2006,
       author = {{Jiang}, W.~Z.},
        title = "{Roles of isoscalar hyperons in probing the density dependence of the nuclear symmetry energy}",
      journal = {Physics Letters B},
         year = 2006,
        month = nov,
       volume = {642},
       number = {1-2},
        pages = {28-33},
          doi = {10.1016/j.physletb.2006.09.020}
}

@article{Yakovlev2004,
  title={Neutron star cooling},
  author={Yakovlev, D. G. and Pethick, C. J.},
  journal={Annual Review of Astronomy and Astrophysics},
  volume={42},
  pages={169--210},
  year={2004},
  publisher={Annual Reviews}
}

@article{Page2006,
  title={The cooling of compact stars},
  author={Page, Dany and Geppert, Ulrich and Weber, Fridolin},
  journal={Nuclear Physics A},
  volume={777},
  pages={497--530},
  year={2006},
  publisher={Elsevier}
}

@article{Andersson1998,
  author = {Andersson, Nils and Kokkotas, Kostas D.},
  title = {Towards gravitational wave asteroseismology for neutron stars},
  journal = {Monthly Notices of the Royal Astronomical Society},
  volume = {299},
  number = {4},
  pages = {1059--1068},
  year = {1998},
  publisher = {Oxford University Press}
}

@article{Lattimer2001,
  author = {Lattimer, J. M. and Prakash, M.},
  title = {Neutron Star Structure and the Equation of State},
  journal = {The Astrophysical Journal},
  volume = {550},
  pages = {426--442},
  year = {2001}
}

@article{Breu2016,
  author = {Breu, Cosima and Rezzolla, Luciano},
  title = {Maximum mass, moment of inertia and compactness of relativistic stars},
  journal = {Monthly Notices of the Royal Astronomical Society},
  volume = {459},
  number = {1},
  pages = {646--656},
  year = {2016},
  publisher = {Oxford University Press}
}

@article{Yagi2013_Science,
  author = {Yagi, Kent and Yunes, Nicol\'as},
  title = {I-Love-Q: Unexpected Universal Relations for Neutron Stars and Quark Stars},
  journal = {Science},
  volume = {341},
  number = {6144},
  pages = {365--368},
  year = {2013},
  doi = {10.1126/science.1236462},
  publisher = {American Association for the Advancement of Science}
}

@article{Buchdahl1959,
  author = {Buchdahl, H. A.},
  title = {General Relativistic Fluid Spheres},
  journal = {Physical Review},
  volume = {116},
  pages = {1027--1034},
  year = {1959}
}

@article{Buchdahl1967,
  author = {Buchdahl, H. A.},
  title = {General Relativistic Fluid Spheres. III. A Static Gaseous Model},
  journal = {The Astrophysical Journal},
  volume = {147},
  pages = {310},
  year = {1967}
}

@article{Tolman1939,
  author = {Tolman, R. C.},
  title = {Static Solutions of Einstein's Field Equations for Spheres of Fluid},
  journal = {Physical Review},
  volume = {55},
  issue = {4},
  pages = {364--373},
  year = {1939},
  publisher = {American Physical Society}
}

@ARTICLE{Lind1992,
       author = {{Lindblom}, Lee},
        title = "{Determining the nuclear equation of state from neutron-star masses and radii}",
      journal = {The Astrophysical Journal},
         year = 1992,
        month = oct,
       volume = {398},
        pages = {569},
          doi = {10.1086/171882},
       adsurl = {https://ui.adsabs.harvard.edu/abs/1992ApJ...398..569L},
}

@article{Chen2015,
  title = {On the compactness of neutron stars},
  author = {Chen, Wei-Chia and Piekarewicz, J.},
  journal = {Physical Review Letters},
  volume = {115},
  issue = {16},
  pages = {161101},
  year = {2015},
  month = {Oct},
  publisher = {American Physical Society},
  doi = {10.1103/PhysRevLett.115.161101},
  url = {https://link.aps.org/doi/10.1103/PhysRevLett.115.161101}
}

@ARTICLE{Mann2006,
       author = {{Mannheim}, Philip D.},
        title = "{Alternatives to dark matter and dark energy}",
      journal = {Progress in Particle and Nuclear Physics},
         year = 2006,
        month = apr,
       volume = {56},
       number = {2},
        pages = {340-445},
          doi = {10.1016/j.ppnp.2005.08.001}
}

@article{Planck2020,
  author = {Aghanim, N. and others},
  collaboration = {Planck Collaboration},
  title = {Planck 2018 results. VI. Cosmological parameters},
  journal = {Astronomy \& Astrophysics},
  volume = {641},
  pages = {A6},
  year = {2020},
  publisher = {EDP Sciences}
}

@article{Kouvaris2008,
  author = {Kouvaris, Chris},
  title = {WIMP annihilation and cooling of neutron stars},
  journal = {Physical Review D},
  volume = {77},
  pages = {023006},
  year = {2008}
}

@article{Brito2015,
  title = {Proca stars: Gravitating Bose-Einstein condensates of massive spin 1 particles},
  author = {Brito, Richard and Cardoso, Vitor and Herdeiro, Carlos A. R. and Radu, Eugen},
  journal = {Physics Letters B},
  volume = {752},
  pages = {291--295},
  year = {2016},
  doi = {10.1016/j.physletb.2015.11.051}
}

@article{Diedrichs2023,
  title = {Tidal deformability of fermion-boson stars: Neutron stars admixed with ultralight dark matter},
  author = {Diedrichs, Robin F. and Becker, Nils and Jockel, C{\'e}dric and Christian, Jan-Erik and Sagunski, Laura and Schaffner-Bielich, J{\"u}rgen},
  journal = {Physical Review D},
  volume = {108},
  number = {6},
  pages = {064009},
  year = {2023},
  doi = {10.1103/PhysRevD.108.064009}
}

@article{Meliani2016,
  title={GR-AMRVAC code applications: accretion onto compact objects, boson stars versus black holes},
  author={Meliani, Z and Grandcl{\'e}ment, P and Casse, F and Vincent, FH and Straub, O and Dauvergne, F},
  journal={Classical and Quantum Gravity},
  volume={33},
  number={15},
  pages={155010},
  year={2016},
  publisher={IOP Publishing}
}

@article{Baym2018,
  title={Testing dark decays of baryons in neutron stars},
  author={Baym, Gordon and Beck, DH and Geltenbort, Peter and Shelton, Jessie},
  journal={Physical review letters},
  volume={121},
  number={6},
  pages={061801},
  year={2018},
  publisher={APS}
}

@article{Berryman2022,
  title={Neutron stars with baryon number violation, probing dark sectors},
  author={Berryman, Jeffrey M and Gardner, Susan and Zakeri, Mohammadreza},
  journal={Symmetry},
  volume={14},
  number={3},
  pages={518},
  year={2022},
  publisher={MDPI}
}

@article{Riley2019,
  title={A NICER View of PSR J0030+0451: Millisecond Pulsar Parameter Estimation},
  author={Riley, Thomas E and others},
  journal={The Astrophysical Journal Letters},
  volume={887},
  number={1},
  pages={L21},
  year={2019},
  doi={10.3847/2041-8213/ab481c},
  publisher={IOP Publishing}
}

@article{Miller2019,
  title={PSR J0030+0451 Mass and Radius from NICER Data and Implications for the Properties of Neutron Star Matter},
  author={Miller, M. C. and others},
  journal={The Astrophysical Journal Letters},
  volume={887},
  number={1},
  pages={L24},
  year={2019},
  doi={10.3847/2041-8213/ab50c5},
  publisher={IOP Publishing}
}

@article{Riley2021,
  title={A NICER View of the Massive Pulsar PSR J0740+6620 Informed by Radio Timing and XMM-Newton Spectroscopy},
  author={Riley, Thomas E. and others},
  journal={The Astrophysical Journal Letters},
  volume={918},
  number={2},
  pages={L27},
  year={2021},
  doi={10.3847/2041-8213/ac0a81},
  publisher={IOP Publishing}
}

@article{Miller2021,
  title={The Radius of PSR J0740+6620 from NICER and XMM-Newton Data},
  author={Miller, M. C. and others},
  journal={The Astrophysical Journal Letters},
  volume={918},
  number={2},
  pages={L28},
  year={2021},
  doi={10.3847/2041-8213/ac089b},
  publisher={IOP Publishing}
}

@article{Goldman1989,
  author = {Goldman, I. and Nussinov, S.},
  title = {Weakly interacting massive particles and neutron stars},
  journal = {Physical Review D},
  volume = {40},
  pages = {3221--3230},
  year = {1989}
}

@article{Rutherford2023,
  author = {Rutherford, N. and Raaijmakers, G. and Prescod-Weinstein, C. and Watts, A.},
  title = {Constraining bosonic asymmetric dark matter with neutron star mass-radius measurements},
  journal = {Physical Review D},
  volume = {107},
  number = {10},
  pages = {103051},
  year = {2023},
  doi = {10.1103/PhysRevD.107.103051}
}

@article{Bertone2008,
  author = {Bertone, G. and Fairbairn, M.},
  title = {Compact Stars as Dark Matter Probes},
  journal = {Physical Review D},
  volume = {77},
  number = {4},
  pages = {043515},
  year = {2008},
  doi = {10.1103/PhysRevD.77.043515}
}

@article{Ellis2018,
  author = {Ellis, J. and H\"utsi, G. and Kannike, K. and Marzola, L. and Raidal, M. and Vaskonen, V.},
  title = {Dark Matter Effects On Neutron Star Properties},
  journal = {Physical Review D},
  volume = {97},
  number = {12},
  pages = {123007},
  year = {2018},
  doi = {10.1103/PhysRevD.97.123007}
}

@article{Shirke2023_JCAP,
doi = {10.1088/1475-7516/2023/12/008},
url = {https://doi.org/10.1088/1475-7516/2023/12/008},
year = {2023},
month = {dec},
publisher = {IOP Publishing},
volume = {2023},
number = {12},
pages = {008},
author = {Shirke, Swarnim and Ghosh, Suprovo and Chatterjee, Debarati and Sagunski, Laura and Schaffner-Bielich, Jürgen},
title = {R-modes as a new probe of dark matter in neutron stars},
journal = {Journal of Cosmology and Astroparticle Physics},
}

@article{Kark2022,
  title = {Bosonic dark matter in neutron stars and its effect on gravitational wave signal},
  author = {Rafiei Karkevandi, Davood and Shakeri, Soroush and Sagun, Violetta and Ivanytskyi, Oleksii},
  journal = {Phys. Rev. D},
  volume = {105},
  issue = {2},
  pages = {023001},
  numpages = {18},
  year = {2022},
  month = {Jan},
  publisher = {American Physical Society},
  doi = {10.1103/PhysRevD.105.023001},
  url = {https://link.aps.org/doi/10.1103/PhysRevD.105.023001}
}

@article{PDG2024,
    author = "Navas, S. and others",
    collaboration = "Particle Data Group",
    title = "{Review of Particle Physics}",
    journal = "Phys. Rev. D",
    volume = "110",
    issue = "3",
    pages = "030001",
    year = "2024",
    doi = "10.1103/PhysRevD.110.030001"
}

@article{Tolos2015,
  author = {Tolos, Laura and Schaffner-Bielich, J\"urgen},
  title = {Dark compact planets},
  journal = {Physical Review D},
  volume = {92},
  number = {12},
  pages = {123002},
  year = {2015},
  doi = {10.1103/PhysRevD.92.123002}
}

@ARTICLE{Cia2011,
       author = {{Ciarcelluti}, Paolo and {Sandin}, Fredrik},
        title = "{Have neutron stars a dark matter core?}",
      journal = {Physics Letters B},
         year = 2011,
        month = jan,
       volume = {695},
       number = {1-4},
        pages = {19-21},
          doi = {10.1016/j.physletb.2010.11.021},
}

@article{jockel2024fermion,
  title={Fermion Proca stars: Vector-dark-matter-admixed neutron stars},
  author={Jockel, C{\'e}dric and Sagunski, Laura},
  journal={Particles},
  volume={7},
  number={1},
  pages={52--79},
  year={2024},
  publisher={MDPI}
}

@article{Liebling2017,
  author  = {Liebling, Steven L. and Palenzuela, Carlos},
  title   = {Boson Stars},
  journal = {Living Reviews in Relativity},
  year    = {2017},
  volume  = {20},
  number  = {1},
  pages   = {5},
  doi     = {10.1007/s41114-017-0007-0},
  url     = {https://doi.org/10.1007/s41114-017-0007-0}
}

@article{Narain2006,
  author = {Narain, G. and Schaffner-Bielich, J. and Mishustin, I. N.},
  title = {Compact stars made of fermionic dark matter and boson stars},
  journal = {Physical Review D},
  volume = {74},
  pages = {063003},
  year = {2006}
}

@ARTICLE{Ivan2020,
       author = {{Ivanytskyi}, O. and {Sagun}, V. and {Lopes}, I.},
        title = "{Neutron stars: New constraints on asymmetric dark matter}",
      journal = {\prd},
         year = {2020},
       volume = {102},
       number = {6},
          eid = {063028},
        pages = {063028},
          doi = {10.1103/PhysRevD.102.063028},
}

@article{Rout2025,
  title={Decoding dark matter admixed neutron stars: From static structure to rotational deformation},
  author={Routaray, Pinku and Chakrawarty, Abirbhav and Patra, NK and Kumar, Bharat},
  journal={Physics of the Dark Universe},
  volume={50},
  pages={102093},
  year={2025},
  doi={10.1016/j.dark.2025.102093},
  publisher={Elsevier}
}

@article{Biesdorf2025,
  title = {Masquerading hybrid stars with dark matter},
  author = {Biesdorf, Carline and Schaffner-Bielich, J{\"u}rgen and Tolos, Laura},
  journal = {Physical Review D},
  volume = {111},
  number = {8},
  pages = {083038},
  year = {2025},
  doi = {10.1103/PhysRevD.111.083038},
  publisher = {American Physical Society}
}

@article{Mukh2016,
  title = {Quark stars admixed with dark matter},
  author = {Mukhopadhyay, Payel and Schaffner-Bielich, J\"urgen},
  journal = {Phys. Rev. D},
  volume = {93},
  issue = {8},
  pages = {083009},
  numpages = {10},
  year = {2016},
  month = {Apr},
  publisher = {American Physical Society},
  doi = {10.1103/PhysRevD.93.083009},
  url = {https://link.aps.org/doi/10.1103/PhysRevD.93.083009}
}

@article{Moira2019,
  title = {Asymmetric dark stars and neutron star stability},
  author = {Gresham, Moira I. and Zurek, Kathryn M.},
  journal = {Phys. Rev. D},
  volume = {99},
  issue = {8},
  pages = {083008},
  numpages = {17},
  year = {2019},
  month = {Apr},
  publisher = {American Physical Society},
  doi = {10.1103/PhysRevD.99.083008},
  url = {https://link.aps.org/doi/10.1103/PhysRevD.99.083008}
}

@article{Das2020,
    author = {Das, Harish Chandra and Kumar, Ankit and Kumar, Bharat and Biswal, S. Kumar and Nakatsukasa, Takashi and Li, Ang and Patra, S. K.},
    title = {Effects of dark matter on the nuclear and neutron star matter},
    journal = {Monthly Notices of the Royal Astronomical Society},
    volume = {495},
    number = {4},
    pages = {4893--4903},
    year = {2020},
    month = {07},
    issn = {0035-8711},
    doi = {10.1093/mnras/staa1435},
    url = {https://doi.org/10.1093/mnras/staa1435},
    publisher = {Oxford University Press}
}

@article{Kain2021,
  title = {Dark matter admixed neutron stars},
  author = {Kain, Ben},
  journal = {Phys. Rev. D},
  volume = {103},
  issue = {4},
  pages = {043009},
  numpages = {12},
  year = {2021},
  month = {Feb},
  publisher = {American Physical Society},
  doi = {10.1103/PhysRevD.103.043009},
  url = {https://link.aps.org/doi/10.1103/PhysRevD.103.043009}
}

@article{Yang2021,
  title = {Strange stars with a mirror-dark-matter core confronting with the observations of compact stars},
  author = {Yang, Shu-Hua and Pi, Chun-Mei and Zheng, Xiao-Ping},
  journal = {Phys. Rev. D},
  volume = {104},
  issue = {8},
  pages = {083016},
  numpages = {9},
  year = {2021},
  month = {Oct},
  publisher = {American Physical Society},
  doi = {10.1103/PhysRevD.104.083016},
  url = {https://link.aps.org/doi/10.1103/PhysRevD.104.083016}
}

@article{Leung2022_Tidal,
  author  = {Leung, Kwing-Lam and Chu, Ming-Chung and Lin, Lap-Ming},
  title   = {Tidal Deformability of Dark Matter Admixed Neutron Stars},
  journal = {Physical Review D},
  volume  = {105},
  number  = {12},
  pages   = {123010},
  year    = {2022},
  doi     = {10.1103/PhysRevD.105.123010},
  url     = {https://doi.org/10.1103/PhysRevD.105.123010},
}

@article{Collier2022,
  title = {Tidal Love numbers of novel and admixed celestial objects},
  author = {Collier, Michael and Croon, Djuna and Leane, Rebecca K.},
  journal = {Phys. Rev. D},
  volume = {106},
  issue = {12},
  pages = {123027},
  numpages = {10},
  year = {2022},
  month = {Dec},
  publisher = {American Physical Society},
  doi = {10.1103/PhysRevD.106.123027},
  url = {https://link.aps.org/doi/10.1103/PhysRevD.106.123027}
}

@article{Komo2022,
  title = {How Perturbative QCD Constrains the Equation of State at Neutron-Star Densities},
  author = {Komoltsev, Oleg and Kurkela, Aleksi},
  journal = {Phys. Rev. Lett.},
  volume = {128},
  issue = {20},
  pages = {202701},
  numpages = {7},
  year = {2022},
  month = {May},
  publisher = {American Physical Society},
  doi = {10.1103/PhysRevLett.128.202701},
  url = {https://link.aps.org/doi/10.1103/PhysRevLett.128.202701}
}

@article{Miao2022,
  author = {Miao, Zhiqiang and Zhu, Yaofeng and Li, Ang and Huang, Feng},
  title = {Dark matter admixed neutron star properties in the light of {X}-ray pulse profile observations},
  journal = {The Astrophysical Journal},
  volume = {936},
  number = {1},
  pages = {69},
  year = {2022},
  doi = {10.3847/1538-4357/ac8544}
}

@article{Deng2022,
  title = {Second Love number of dark compact planets and neutron stars with dark matter},
  author = {Dengler, Yannick and Schaffner-Bielich, J\"urgen and Tolos, Laura},
  journal = {Phys. Rev. D},
  volume = {105},
  issue = {4},
  pages = {043013},
  numpages = {10},
  year = {2022},
  month = {Feb},
  publisher = {American Physical Society},
  doi = {10.1103/PhysRevD.105.043013},
  url = {https://link.aps.org/doi/10.1103/PhysRevD.105.043013}
}

@article{Shak2024,
  title = {Bosonic dark matter in light of the NICER precise mass-radius measurements},
  author = {Shakeri, Soroush and Karkevandi, Davood Rafiei},
  journal = {Phys. Rev. D},
  volume = {109},
  issue = {4},
  pages = {043029},
  numpages = {23},
  year = {2024},
  month = {Feb},
  publisher = {American Physical Society},
  doi = {10.1103/PhysRevD.109.043029},
  url = {https://link.aps.org/doi/10.1103/PhysRevD.109.043029}
}

@article{Sun2024,
  title = {New criterion for the existence of dark matter in neutron stars},
  author = {Sun, Hongyi and Wen, Dehua},
  journal = {Phys. Rev. D},
  volume = {109},
  issue = {12},
  pages = {123037},
  numpages = {12},
  year = {2024},
  month = {Jun},
  publisher = {American Physical Society},
  doi = {10.1103/PhysRevD.109.123037},
  url = {https://link.aps.org/doi/10.1103/PhysRevD.109.123037}
}

@article{Hong2024,
  title = {Mixed dark matter models for the peculiar compact object in remnant HESS $\mathrm{J}1731\ensuremath{-}347$ and their implications for gravitational wave properties},
  author = {Hong, Bin and Ren, ZhongZhou},
  journal = {Phys. Rev. D},
  volume = {109},
  issue = {2},
  pages = {023002},
  numpages = {18},
  year = {2024},
  month = {Jan},
  publisher = {American Physical Society},
  doi = {10.1103/PhysRevD.109.023002},
  url = {https://link.aps.org/doi/10.1103/PhysRevD.109.023002}
}

@article{Liu2024,
  title = {Effects from dark matter halos on x-ray pulsar pulse profiles},
  author = {Liu, Yukun and Li, Hong-Bo and Gao, Yong and Shao, Lijing and Hu, Zexin},
  journal = {Phys. Rev. D},
  volume = {110},
  issue = {8},
  pages = {083018},
  numpages = {10},
  year = {2024},
  month = {Oct},
  publisher = {American Physical Society},
  doi = {10.1103/PhysRevD.110.083018},
  url = {https://link.aps.org/doi/10.1103/PhysRevD.110.083018}
}

@article{Yang2025,
  title = {Strange stars admixed with mirror dark matter: Confronting observations of XTE J1814-338},
  author = {Yang, Shu-Hua and Pi, Chun-Mei and Weber, Fridolin},
  journal = {Phys. Rev. D},
  volume = {111},
  issue = {4},
  pages = {043037},
  numpages = {9},
  year = {2025},
  month = {Feb},
  publisher = {American Physical Society},
  doi = {10.1103/PhysRevD.111.043037},
  url = {https://link.aps.org/doi/10.1103/PhysRevD.111.043037}
}

@article{Zhou2026,
  title = {Stability Boundary of Neutron-Dark Matter Mixed Stars},
  author = {Zhou, Xiao-Ding and Chen, Tian-Shun and Wu, Si-Man and Zhang, Kilar},
  journal = {Chinese Physics C},
  year = {2026},
  doi = {10.1088/1674-1137/ae86dd},
  url = {https://doi.org/10.1088/1674-1137/ae86dd}
}

@article{Issi2026,
  title = {Dark matter heating in evolving protoneutron stars: A two-fluid approach},
  author = {Issifu, Adamu and Thakur, Prashant and Karkevandi, Davood Rafiei and da Silva, Franciele M. and Menezes, D\'ebora P. and Lim, Y. and Frederico, Tobias},
  journal = {Phys. Rev. D},
  volume = {113},
  issue = {10},
  pages = {103042},
  numpages = {16},
  year = {2026},
  month = {May},
  publisher = {American Physical Society},
  doi = {10.1103/bhm3-jzjq},
  url = {https://link.aps.org/doi/10.1103/bhm3-jzjq}
}

@article{Glendenning1991,
  author = {Glendenning, N. K. and Moszkowski, S. A.},
  title = {Reconciliation of neutron-star masses and binding of the Lambda in hypernuclei},
  journal = {Physical Review Letters},
  volume = {67},
  pages = {2414--2417},
  year = {1991},
  doi = {10.1103/PhysRevLett.67.2414}
}

@article{Long2004,
  author = {Long, W.-H. and Meng, J. and Giai, N. Van and Zhou, S.-G.},
  title = {New effective interactions in relativistic mean field theory with nonlinear terms and density-dependent meson-nucleon coupling},
  journal = {Physical Review C},
  volume = {69},
  pages = {034319},
  year = {2004},
  doi = {10.1103/PhysRevC.69.034319}
}

@article{Typel1999,
  author = {Typel, S. and Wolter, H. H.},
  title = {Relativistic mean field calculations with density-dependent meson-nucleon coupling},
  journal = {Nuclear Physics A},
  volume = {656},
  pages = {331--364},
  year = {1999},
  doi = {10.1016/S0375-9474(99)00310-3}
}

@article{ChenPiekarewicz2014,
  author = {Chen, Wei-Chia and Piekarewicz, J.},
  title = {Building relativistic mean-field models for finite nuclei and neutron stars},
  journal = {Physical Review C},
  volume = {90},
  pages = {044305},
  year = {2014},
  doi = {10.1103/PhysRevC.90.044305}
}

@article{Jiang2007_SLC,
  author = {Jiang, Wei-Zhou and Li, Bao-An and Chen, Lie-Wen},
  title = {Equation of state of isospin-asymmetric nuclear matter in relativistic mean-field models with chiral limits},
  journal = {Physics Letters B},
  volume = {653},
  number = {2--4},
  pages = {184--189},
  year = {2007},
  doi = {10.1016/j.physletb.2007.08.004}
}

@ARTICLE{Jiang2012,
       author = {{Jiang}, Wei-Zhou and {Li}, Bao-An and {Chen}, Lie-Wen},
        title = "{Large-mass neutron stars with hyperonization}",
      journal = {The Astrophysical Journal},
         year = 2012,
        month = sep,
       volume = {756},
       number = {1},
          eid = {56},
        pages = {56},
          doi = {10.1088/0004-637X/756/1/56},
}

@ARTICLE{Ye2025,
       author = {{Ye}, Jing and {Jiang}, Wei-Zhou},
        title = "{Thermal properties of zero sound in asymmetrical nuclear matter}",
      journal = {Physical Review C},
         year = 2025,
        month = may,
       volume = {111},
       number = {5},
          eid = {054315},
        pages = {054315},
          doi = {10.1103/PhysRevC.111.054315},
}

@article{Typel2010,
  author = {Typel, S. and R{\"o}pke, G. and Kl{\"a}hn, T. and Blaschke, D. and Wolter, H. H.},
  title = {Composition and thermodynamics of nuclear matter with light clusters},
  journal = {Physical Review C},
  volume = {81},
  pages = {015803},
  year = {2010},
  doi = {10.1103/PhysRevC.81.015803}
}

@article{Grill2014,
  author = {Grill, Fabrizio and Pais, Helena and Providencia, Constanca and Vidana, Isaac and Avancini, Sidney S.},
  title = {Equation of state and thickness of the inner crust of neutron stars},
  journal = {Physical Review C},
  volume = {90},
  number = {4},
  pages = {045803},
  year = {2014},
  doi = {10.1103/PhysRevC.90.045803}
}

@article{Scurto2024,
  author = {Scurto, Luigi and Pais, Helena and Gulminelli, Francesca},
  title = {General predictions of neutron star properties using unified relativistic mean-field equations of state},
  journal = {Physical Review D},
  volume = {109},
  number = {10},
  pages = {103015},
  year = {2024},
  doi = {10.1103/PhysRevD.109.103015},
}

@article{Maruyama2005,
  author = {Maruyama, Toshiki and Tatsumi, Toshitaka and Voskresensky, Dmitri N. and Tanigawa, Tomonori and Chiba, Satoshi},
  title = {Nuclear pasta structures and the charge screening effect},
  journal = {Physical Review C},
  volume = {72},
  number = {1},
  pages = {015802},
  year = {2005},
  doi = {10.1103/PhysRevC.72.015802}
}

@article{Taninah2020,
  author = {Taninah, A. and Agbemava, S. E. and Afanasjev, A. V. and Ring, P.},
  title = {Parametric correlations in energy density functionals},
  journal = {Physics Letters B},
  volume = {800},
  pages = {135065},
  year = {2020},
  doi = {10.1016/j.physletb.2019.135065}
}

@article{Goriely2013,
  author = {Goriely, S. and Chamel, N. and Pearson, J. M.},
  title = {Further explorations of {Skyrme}-{Hartree}-{Fock}-{Bogoliubov} mass formulas. {XIII}. The 2012 atomic mass evaluation and the symmetry coefficient},
  journal = {Physical Review C},
  volume = {88},
  pages = {024308},
  year = {2013},
  doi = {10.1103/PhysRevC.88.024308}
}

@article{Pearson2018,
  author = {Pearson, J. M. and Chamel, N. and Potekhin, A. Y. and Fantina, A. F. and Ducoin, C. and Dutta, A. K. and Goriely, S.},
  title = {Unified equations of state for cold non-accreting neutron stars with Brussels--Montreal functionals. I. Role of symmetry energy},
  journal = {Monthly Notices of the Royal Astronomical Society},
  volume = {481},
  pages = {2994--3026},
  year = {2018},
  doi = {10.1093/mnras/sty2413}
}

@article{Grams2022,
  author = {Grams, G. and Margueron, J. and Somasundaram, R. and Reddy, S.},
  title = {Confronting a set of {Skyrme} and {chiEFT} predictions for the crust of neutron stars: on the origin of uncertainties in model predictions},
  journal = {European Physical Journal A},
  volume = {58},
  pages = {56},
  year = {2022},
  doi = {10.1140/epja/s10050-022-00706-w}
}

@article{DouchinHaensel2001,
  author = {Douchin, F. and Haensel, P.},
  title = {A unified equation of state of dense matter and neutron star structure},
  journal = {Astronomy and Astrophysics},
  volume = {380},
  pages = {151--167},
  year = {2001},
  doi = {10.1051/0004-6361:20011402}
}

@article{Bombaci2018,
  author = {Bombaci, I. and Logoteta, D.},
  title = {Equation of state of dense nuclear matter and neutron star structure from nuclear chiral interactions},
  journal = {Astronomy and Astrophysics},
  volume = {609},
  pages = {A128},
  year = {2018},
  doi = {10.1051/0004-6361/201731604}
}

@article{Typel2015_CompOSE,
  author = {Typel, S. and Oertel, M. and Kl{\"a}hn, T.},
  title = {CompOSE CompStar online supernova equations of state harmonising the concert of nuclear physics and astrophysics},
  journal = {Physics of Particles and Nuclei},
  volume = {46},
  pages = {633--664},
  year = {2015},
  doi = {10.1134/S1063779615040061}
}

@article{Typel2022_CompOSE,
  author = {Typel, S. and Oertel, M. and Kl{\"a}hn, T. and others},
  title = {CompOSE reference manual},
  journal = {European Physical Journal A},
  volume = {58},
  pages = {221},
  year = {2022},
  doi = {10.1140/epja/s10050-022-00847-y},
}

@article{Choudhury2024_J0437,
  title={A NICER View of the Nearest and Brightest Millisecond Pulsar: PSR J0437-4715},
  author={Choudhury, Devarshi and others},
  journal={The Astrophysical Journal Letters},
  volume={971},
  number={1},
  pages={L20},
  year={2024},
  doi={10.3847/2041-8213/ad5a6f},
  publisher={IOP Publishing}
}

@article{Doroshenko2022_HESS,
  title={A strangely light neutron star within a supernova remnant},
  author={Doroshenko, Victor and Suleimanov, Valery and Puehlhofer, Gerd and Santangelo, Andrea},
  journal={Nature Astronomy},
  volume={6},
  pages={1444--1451},
  year={2022},
  doi={10.1038/s41550-022-01800-1},
  publisher={Springer Nature}
}

@misc{Doroshenko2022_HESS_Zenodo,
  author = {Doroshenko, Victor and Suleimanov, Valery and Puehlhofer, Gerd and Santangelo, Andrea},
  title = {Data for: A strangely light neutron star within a supernova remnant},
  year = {2022},
  publisher = {Zenodo},
  doi = {10.5281/zenodo.6702216},
  url = {https://zenodo.org/records/6702216}
}

@article{BPS1971,
  title={The ground state of matter at high densities: equation of state and stellar models},
  author={Baym, Gordon and Pethick, Christopher and Sutherland, Peter},
  journal={The Astrophysical Journal},
  volume={170},
  pages={299--317},
  year={1971},
  doi={10.1086/151216}
}

@article{Iida1997,
  title={Spin down of neutron stars and compositional transitions in the cold crustal matter},
  author={Iida, Kei and Sato, Katsuhiko},
  journal={The Astrophysical Journal},
  volume={477},
  number={1},
  pages={294--312},
  year={1997}
}

@article{Aprile2023_XENONnT,
  author = {Aprile, E. and others},
  collaboration = {XENON Collaboration},
  title = {First Dark Matter Search with Nuclear Recoils from the XENONnT Experiment},
  journal = {Physical Review Letters},
  volume = {131},
  issue = {4},
  pages = {041003},
  year = {2023},
  doi = {10.1103/PhysRevLett.131.041003},
  publisher = {American Physical Society}
}

@article{Aalbers2022_LZ,
  author = {Aalbers, J. and others},
  collaboration = {LZ Collaboration},
  title = {First Dark Matter Search Results from the LUX-ZEPLIN (LZ) Experiment},
  journal = {Physical Review Letters},
  volume = {129},
  issue = {16},
  pages = {161801},
  year = {2022},
  doi = {10.1103/PhysRevLett.129.161801},
  publisher = {American Physical Society}
}

@article{Bringmann2019_CRDM,
  author = {Bringmann, Torsten and Pospelov, Maxim},
  title = {Novel Direct Detection Constraints on Light Dark Matter},
  journal = {Physical Review Letters},
  volume = {122},
  issue = {17},
  pages = {171801},
  year = {2019},
  doi = {10.1103/PhysRevLett.122.171801},
  publisher = {American Physical Society}
}

@article{FinchSkea1989,
  author = {Finch, M. R. and Skea, J. E. F.},
  title = {A realistic stellar model based on an ansatz of Duorah and Ray},
  journal = {Classical and Quantum Gravity},
  volume = {6},
  pages = {467--476},
  year = {1989}
}

@article{ji2025equation,
  title={The Equation of State of Neutron Stars: Theoretical Models, Observational Constraints, and Future Perspectives},
  author={Ji, Zuhua and Chen, Jiarui},
  journal={arXiv preprint},
  year={2025},
  eprint={2502.05513},
  archivePrefix={arXiv},
  primaryClass={astro-ph.HE}
}

@article{Baldo2016,
  author = {Baldo, M. and Burgio, G. F.},
  title = {The nuclear symmetry energy},
  journal = {Progress in Particle and Nuclear Physics},
  volume = {91},
  pages = {203--258},
  year = {2016},
  doi = {10.1016/j.ppnp.2016.06.006},
  url = {https://www.sciencedirect.com/science/article/pii/S0146641016300254},
 }

@article{Li2018,
  title = {Nucleon effective masses in neutron-rich matter},
  author = {Li, Bao-An and Cai, Bao-Jun and Chen, Lie-Wen and Xu, Jun},
  journal = {Progress in Particle and Nuclear Physics},
  volume = {99},
  pages = {29--119},
  year = {2018},
  issn = {0146-6410},
  doi = {10.1016/j.ppnp.2018.01.001},
  url = {https://doi.org/10.1016/j.ppnp.2018.01.001},
  publisher = {Elsevier}
}

@article{Grippa2025,
  author = {Grippa, Francesco and Lambiase, Gaetano and Poddar, Tanmay Kumar},
  title = {Searching for New Physics in an Ultradense Environment: A Review on Dark Matter Admixed Neutron Stars},
  journal = {Universe},
  volume = {11},
  number = {3},
  pages = {74},
  year = {2025},
  doi = {10.3390/universe11030074},
}

@article{Kumar2024,
  author = {Kumar, Ankit and Sotani, Hajime},
  title = {Constraints on the parameter space in dark matter admixed neutron stars},
  journal = {Physical Review D},
  volume = {110},
  number = {6},
  pages = {063001},
  year = {2024},
  doi = {10.1103/PhysRevD.110.063001}
}

@article{Lattimer2014,
  title={Constraints on the symmetry energy using the mass-radius relation of neutron stars},
  author={Lattimer, James M and Steiner, Andrew W},
  journal={The European Physical Journal A},
  volume={50},
  number={2},
  pages={40},
  year={2014},
  publisher={Springer}
}

@ARTICLE{Fatt2018,
       author = {{Fattoyev}, F.~J. and {Piekarewicz}, J. and {Horowitz}, C.~J.},
        title = "{Neutron Skins and Neutron Stars in the Multimessenger Era}",
      journal = {Physical Review Letters},
         year = 2018,
        month = apr,
       volume = {120},
       number = {17},
          eid = {172702},
        pages = {172702},
          doi = {10.1103/PhysRevLett.120.172702},
       adsurl = {https://ui.adsabs.harvard.edu/abs/2018PhRvL.120q2702F},
}

@article{Li2021,
  title={Progress in constraining nuclear symmetry energy using neutron star observables since GW170817},
  author={Li, Bao-An and Cai, Bao-Jun and Xie, Wen-Jie and Zhang, Nai-Bo},
  journal={Universe},
  volume={7},
  number={6},
  pages={182},
  year={2021},
  publisher={MDPI}
}

@ARTICLE{Adh2021,
       author = {{Adhikari}, D. and {Albayrak}, H. and {Androic}, D. and {Aniol}, K. and {Armstrong}, D.~S. and {Arrington}, J. and {Atac}, H. and {Averett}, T. and {Barkey}, C. and {Bellini}, V. and et al.},
        title = "{Accurate Determination of the Neutron Skin Thickness of 208Pb through Parity-Violation in Electron Scattering}",
      journal = {Physical Review Letters},
         year = 2021,
        month = apr,
       volume = {126},
       number = {17},
          eid = {172502},
        pages = {172502},
          doi = {10.1103/PhysRevLett.126.172502},
       adsurl = {https://ui.adsabs.harvard.edu/abs/2021PhRvL.126q2502A},
}

@ARTICLE{Reed2021,
       author = {{Reed}, Brendan T. and {Fattoyev}, F.~J. and {Horowitz}, C.~J. and {Piekarewicz}, J.},
        title = "{Implications of PREX-2 on the Equation of State of Neutron-Rich Matter}",
      journal = {Physical Review Letters},
         year = 2021,
        month = apr,
       volume = {126},
       number = {17},
          eid = {172503},
        pages = {172503},
          doi = {10.1103/PhysRevLett.126.172503},
       adsurl = {https://ui.adsabs.harvard.edu/abs/2021PhRvL.126q2503R},
}

@ARTICLE{Guillot2014,
       author = {Guillot, Sebastien and Rutledge, Robert E.},
        title = "{Rejecting Proposed Dense Matter Equations of State with Quiescent Low-mass X-Ray Binaries}",
      journal = {\apj Letters},
         year = {2014},
       volume = {796},
       number = {1},
        pages = {L3},
          doi = {10.1088/2041-8205/796/1/L3},
}

@ARTICLE{Ozel2016,
       author = {{\"O}zel, Feryal and Psaltis, Dimitrios and G{\"u}ver, Tolga and Baym, Gordon and Heinke, Craig and Guillot, Sebastien},
        title = "{The Dense Matter Equation of State from Neutron Star Radius and Mass Measurements}",
      journal = {\apj},
         year = {2016},
       volume = {820},
       number = {1},
        pages = {28},
          doi = {10.3847/0004-637X/820/1/28},
}

@article{Giangrandi2023,
  author = {Giangrandi, Edoardo and Sagun, Violetta and Ivanytskyi, Oleksii and Provid{\^e}ncia, Constan{\c{c}}a and Dietrich, Tim},
  title = {The Effects of Self-interacting Bosonic Dark Matter on Neutron Star Properties},
  journal = {The Astrophysical Journal},
  volume = {953},
  number = {2},
  pages = {115},
  year = {2023},
  doi = {10.3847/1538-4357/ace104},
}

@article{Das2020_DANS_GW,
  author = {Das, Arpan and Malik, Tuhin and Nayak, Alekha C.},
  title = {Dark matter admixed neutron star properties in the light of gravitational wave observations: a two fluid approach},
  journal = {Physical Review D},
  volume = {105},
  number = {12},
  pages = {123034},
  year = {2022},
  doi = {10.1103/PhysRevD.105.123034},
}
\end{document}